\documentclass[journal]{IEEEtran}
\usepackage{amsmath,amsfonts}
\usepackage{algorithm}
\usepackage{textcomp}
\usepackage{stfloats}
\usepackage{url}
\usepackage{verbatim}
\usepackage{graphicx}
\usepackage{epstopdf}
\usepackage{cite}
\usepackage{algpseudocode}
\usepackage{cite}
\usepackage{graphicx}
\usepackage{subfigure} 
\usepackage{amssymb}
\usepackage{svg}
\usepackage{xcolor}
\usepackage{booktabs}
\usepackage{tabularx}
\usepackage{array}
\usepackage{algpseudocode}
\usepackage{multirow}
\usepackage{makecell}
\usepackage{diagbox}
\usepackage{float}
\begin{document}

\title{RSU-Assisted Resource Allocation for Collaborative Perception}

\author{Guowei Liu,
       	Le Liang,~\IEEEmembership{Member,~IEEE,}
       	Chongtao Guo,~\IEEEmembership{Member,~IEEE,}
       	Hao Ye,~\IEEEmembership{Member,~IEEE,}\\
       	and Shi Jin, ~\IEEEmembership{Fellow,~IEEE}
\thanks{Guowei~Liu, Le Liang and Shi Jin are with the School of Information Science and Engineering, Southeast University, Nanjing 210096, China (e-mail: \{grownliu, lliang, jinshi\}@seu.edu.cn). Le Liang is also with the Purple Mountain Laboratories, Nanjing 211111, China.
	
	Hao Ye is with the Department of Electrical and Computer Engineering, University of California, Santa Cruz, CA 95064, USA (e-mail: yehao@ucsc.edu).
	
	Chongtao Guo is with the College of Electronics and Information
	Engineering, Shenzhen University, Shenzhen 518060, China (e-mail: ctguo@szu.edu.cn).}
}



\maketitle

\begin{abstract}

As a pivotal technology for autonomous driving, collaborative perception enables vehicular agents to exchange perceptual data through vehicle-to-everything (V2X) communications, thereby enhancing perception accuracy of all collaborators.
However, existing collaborative perception frameworks often assume ample communication resources, which is usually impractical in real-world vehicular networks.
To address this challenge, this paper investigates the problem of communication resource allocation for collaborative perception and proposes RACooper, a novel RSU-assisted resource allocation framework that maximizes perception accuracy under constrained communication resources. RACooper leverages a hierarchical reinforcement learning model to dynamically allocate communication resources while accounting for real-time sensing data and channel dynamics induced by vehicular mobility. By jointly optimizing spatial confidence metrics and channel state information, our approach ensures efficient feature transmission, enhancing the effectiveness of collaborative perception. Simulation results demonstrate that compared to conventional baseline algorithms, RACooper achieves significant improvements in perception accuracy, especially under bandwidth-constrained scenarios.
\end{abstract}

\IEEEpeerreviewmaketitle

\section{Introduction}

The safety of autonomous driving relies on precise perception of the surrounding environment, enabling vehicles to proactively plan routes to avoid potential risks on the road\cite{ren2022collaborative, li2022v2x}. While current connected autonomous vehicles (CAVs) can perform certain L2 or L3 autonomous driving functions with their own perception systems, standalone perception systems still encounter significant challenges in achieving higher levels of autonomy due to limited perception range and unpredictable occluding objects. These inherent limitations pose significant challenges to the safety and reliability of autonomous driving, and existing studies indicate that the hardware and algorithms of standalone perception systems have reached a point of diminishing returns, with high costs and minimal gains in perception accuracy\cite{noor2020survey}. To address these issues, collaborative perception has been developed as one of promising solutions. Collaborative perception systems utilize vehicle-to-everything (V2X) communication \cite{gholmieh2021c} to encourage CAVs to exchange perception information and integrate information with other CAVs, roadside units (RSUs), or other infrastructures, aiming to significantly improve the overall accuracy and reliability of the perception system.

Depending on the stage of the data processing pipeline at which information is shared, collaborative perception can be classified into three typical paradigms: early collaboration, intermediate collaboration, and late collaboration. In the early collaboration framework, raw sensor data, such as LiDAR point clouds and RGB images, are transmitted via V2X communication before any local processing\cite{chen2019cooper, arnold2020cooperative, cao2025task}. However, the real-time transmission of raw sensor data demands substantial communication resources, and the associated data transmission rate often exceeds network's capacity, constituting the primary bottleneck limiting the practical deployment of early collaboration. Late collaboration, by contrast, involves the exchange of locally processed perception results, such as object detection boxes and semantic segmentation maps\cite{miller2020cooperative, song2023cooperative, yu2022dair}. Although this approach reduces communication load significantly, its performance is still sensitive to the positioning accuracy of vehicles and the reliability of data transmission. While fusion itself may encounter alignment errors due to positioning issues, the main challenge in late collaboration is that each vehicle initially performs object detection independently, failing to make full use of the effective information from collaborative vehicles, thus resulting in suboptimal fusion outcomes.

Intermediate collaboration strikes a balance between communication requirements and perception performance by enabling collaborators to exchange deep feature tensors extracted from the intermediate layers of their local perception models through V2X communication links\cite{chen2019f,guo2021coff, abdel2021v2v, zhou2022multi, xu2022opv2v}. 
Unlike early collaboration, which preserves geometric and semantic detail at the cost of high communication overhead, and late collaboration, which minimizes overhead but sacrifices detail, intermediate collaboration employs data compression and link pruning to reduce communication load while retaining essential features, thereby boosting downstream task performance. Its balanced trade-off between information richness and communication efficiency has made it a key research focus in collaborative perception systems. In the early research on intermediate collaboration, much of the work mainly focused on the perception gains brought by vehicle-to-vehicle (V2V) communication. As the first proposed intermediate collaboration model for collaborative perception, F-Cooper introduces a scheme that operates on features from different representational levels, ranging from low-level voxels to deep spatial features\cite{chen2019f}. This scheme utilizes an element-wise maximum output to fuse the features within overlapping regions. The less similar and more distant neighboring features are, the greater the complementary information they intuitively contribute. This feature complementarity enhances wireless transmission efficiency by enabling the system to prioritize and transmit the most relevant and diverse information, thus reducing redundant data and improving bandwidth utilization.

The aforementioned methods focus on V2V collaboration. However, in real-world scenarios, the quality of collaborative features on CAVs is often limited by the cost of sensors. Compared to V2V collaboration, vehicle-to-infrastructure (V2I) collaboration allows RSUs to provide more stable collaborative information, higher communication rates, and more CAV connections by utilizing better hardware\cite{xu2022v2x}. For instance, in contrast to sensors on vehicles, RSUs offer greater flexibility in sensor placement and can deploy advanced sensors with higher performance. As a result, V2I collaboration can take advantage of the better field of view at the RSU to gather more comprehensive perception information\cite{hu2022where2comm, wang2023vimi}. As the first transformer architecture in collaborative perception, V2X-ViT considered both V2V and V2I collaboration\cite{xu2022v2x}. It introduced a heterogeneous multi-agent attention module to learn the distinct relationships between V2V and V2I. Where2comm proposed spatial confidence maps to reflect the spatial heterogeneity of perception information\cite{hu2022where2comm}, enabling CAVs to share only spatially sparse but crucial perceptual information through V2I links. By prioritizing essential perceptual information, this mechanism enhances overall perception accuracy while making better use of limited communication resource. Meanwhile, a crucial aspect is that RSU-based V2I collaboration is natively compatible with the cellular V2X (C-V2X) architecture, inherently supporting centralized collaboration methods \cite{liu2025deep, sheng2024semantic1}. This allows dynamic allocation of communication and computational resources, which is vital for maintaining quality of service (QoS) and avoiding channel congestion in dense traffic environments.

A key aspect often overlooked in prior work is the practical limitation of communication resources \cite{clancy2024wireless, machardy2018v2x}. For example, the C-V2X standard allocates only 30 MHz of bandwidth in the 5.9 GHz band for V2X services \cite{5gaa2021, chang2024interoperable}. This limited capacity is insufficient for real-time transmission of raw point clouds or high-dimensional features, particularly when shared with other critical safety applications, such as V2V messaging and emergency vehicle warnings. As a result, how to efficiently allocate wireless resources to enhance the perception accuracy of collaborative participants has become a crucial issue\cite{chan2024rsu, abdel2021vehicular, luo2023edgecooper, ye2023accuracy, jia2024c, sheng2024semantic}. Collaborative perception oriented resource allocation for V2I links has been investigated in \cite{chan2024rsu}, which leveraged a two-dimensional perception model to represent coverage energy efficiency. In \cite{abdel2021vehicular}, a reinforcement learning (RL) model is employed to optimize vehicle association and resource allocation at the RSU, with the goal of maximizing vehicles’ satisfaction with the received information. Additionally, due to the inherent limitations of late collaboration, several studies have focused on early collaboration as an alternative to address these challenges \cite{luo2023edgecooper, ye2023accuracy}. Specifically, EdgeCooper employed a voxel-based strategy to establish the relationship between communication and detection \cite{luo2023edgecooper}. It utilizesd multi-hop communication to minimize redundant raw data, thereby enhancing perception accuracy and extending the perception range in RSU. Similar to EdgeCooper, the method in \cite{ye2023accuracy} improves perception accuracy between CAVs and RSUs through early collaboration and computation resource allocation.

Existing studies on resource allocation for collaborative perception have two key limitations. First, they primarily focus on early or late collaboration, neglecting the potential of intermediate feature fusion. Second, and more critically, they operate on a slow timescale tied to sensor sampling rates, which is inadequate to handle the rapid channel fluctuations caused by vehicle mobility and small-scale fading. These dynamics require faster and more adaptive resource management. In response, our research addresses V2I resource allocation for intermediate collaborative perception. We propose RACooper, which allocates communication resources based on feature importance and channel state information (CSI). RACooper employs a hierarchical RL-based algorithm to efficiently allocate resources for optimized perception performance. The main contributions of this paper are summarized as follows:

\begin{itemize}
	\item We formulate the joint problem of feature selection and resource allocation in V2I-assisted collaborative perception and propose the first resource allocation framework for intermediate collaborative perception. Considering the non-analytic nature of the optimization objective, we transform the goal into a more achievable solution, striking an efficient balance between computational complexity and accuracy.
	
	\item We utilize spatial confidence maps to select more important data for transmission, optimizing resource allocation and enhancing collaborative perception. To effectively manage the dynamic interplay between these perceptual priorities and the finite communication resources, we develop a hierarchical RL-based V2I resource allocation algorithm, where power control and resource block (RB) allocation are jointly taken into account.
	
	\item Experimental results demonstrate that, with appropriate reward design and training mechanisms, RACooper can learn from interactions with the CAVs communication environment and identify effective strategies for centralized resource allocation, thereby supporting precise collaborative perception.
	
 \end{itemize}

The rest of the paper is organized as follows. In Section II, we introduce the framework of RACooper. Then Section III formulates the collaborative perception problem of resource allocation in RACooper and transform the non-analytic objective into a sub-optimal solution with analytic expression. Afterwards, we propose a hierarchical RL algorithm in Section IV to solve the optimization problem in RACooper. Finally, simulation results are shown in Section V and the conclusion is drawn in Section VI.

\section{RACooper Framework}

\begin{figure}[t]
	\centering
	\includegraphics[width=\linewidth]{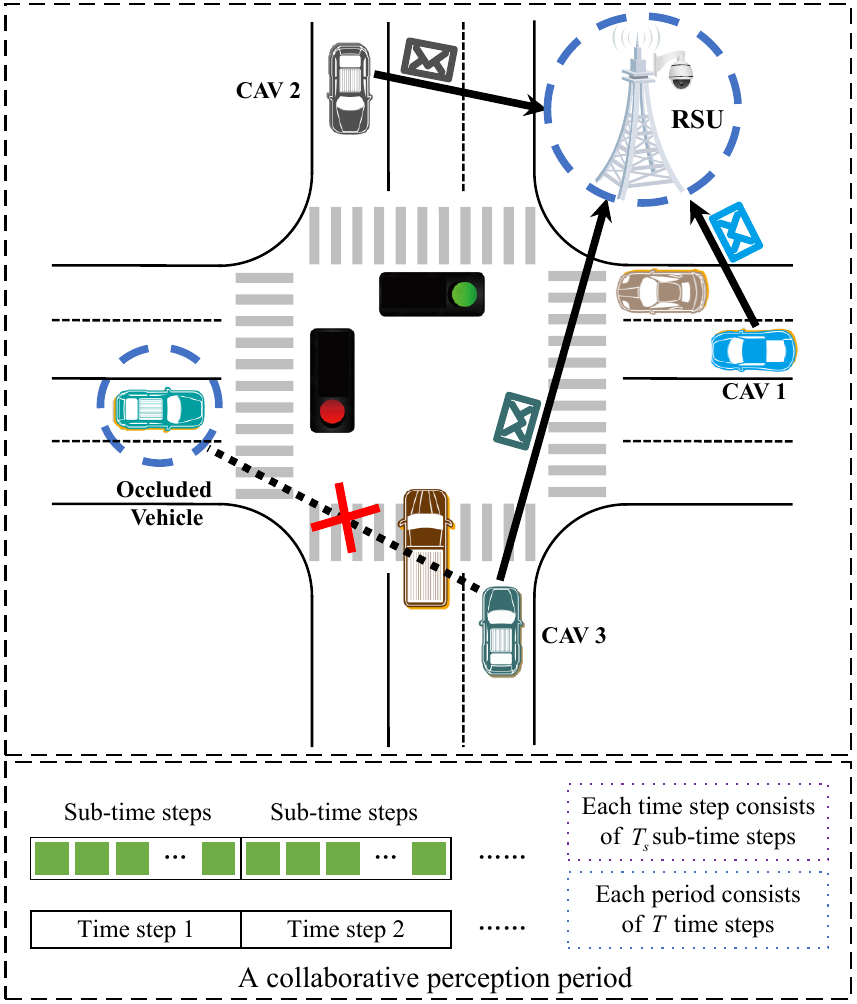}
	\caption{An illustrative scenario of RSU-assisted V2I collaborative perception and the workflow of RACooper.}
	\label{fig_sys}
\end{figure}

As illustrated in Fig. \ref{fig_sys}, we consider an RSU-assisted collaborative perception scenario in this paper, where the RSU collects perception features from CAVs and performs intermediate feature fusion with its own perception features to obtain more accurate and complete perception results. Our research focuses on a centralized resource allocation scheme in C-V2X, where RSU allocates communication resources for CAVs\cite{molina2017lte}. Specifically, each CAV and the RSU are equipped with LiDAR to collect raw perception data. The RSU serves as a central node to gather features uploaded by CAVs within its communication range for fusion. The C-V2X vehicular network connects the CAVs to the RSU via PC5 interfaces. We consider an RSU-assisted collaborative perception system, composed of one RSU and $M$ CAVs. Both the RSU and the CAVs are equipped with single-antenna transceivers.  Consequently, each of the $M$ CAVs establishes a unique V2I link to the RSU. We denote the set of these CAVs and their corresponding V2I links by $\mathcal{M}=\{1,\cdots,M\}$.

The workflow of RACooper, depicted in Fig. \ref{fig_sys}, is structured to reflect the collaborative perception cycle under real-world conditions. We assume that the processing cycle of collaborative perception is synchronized with the sampling period of the RSU and CAV's onboard LiDAR. Accordingly, within each sampling period, the RSU performs several rounds of resource allocation in response to the varying channel conditions. To prevent redundant transmissions from multiple CAVs, it is advantageous to regularly monitor which features have already been received at the RSU and adjust feature selection dynamically. Thus, each collaborative perception period is divided into $T$ steps, with RSU calculating each CAV’s current feature complementarity at the beginning of each step and making resource allocation decisions. Furthermore, due to fast changing channel condition caused by vehicle mobility, each step is considered to consist of $T_s$ sub‑time steps, each corresponding to the channel coherence time. Specifically, within each sub‑time step, small‑scale fading is considered stationary, but it varies independently between sub‑time steps.

\begin{figure*}[htb]
	\centering
	\includegraphics[width=0.95\linewidth]{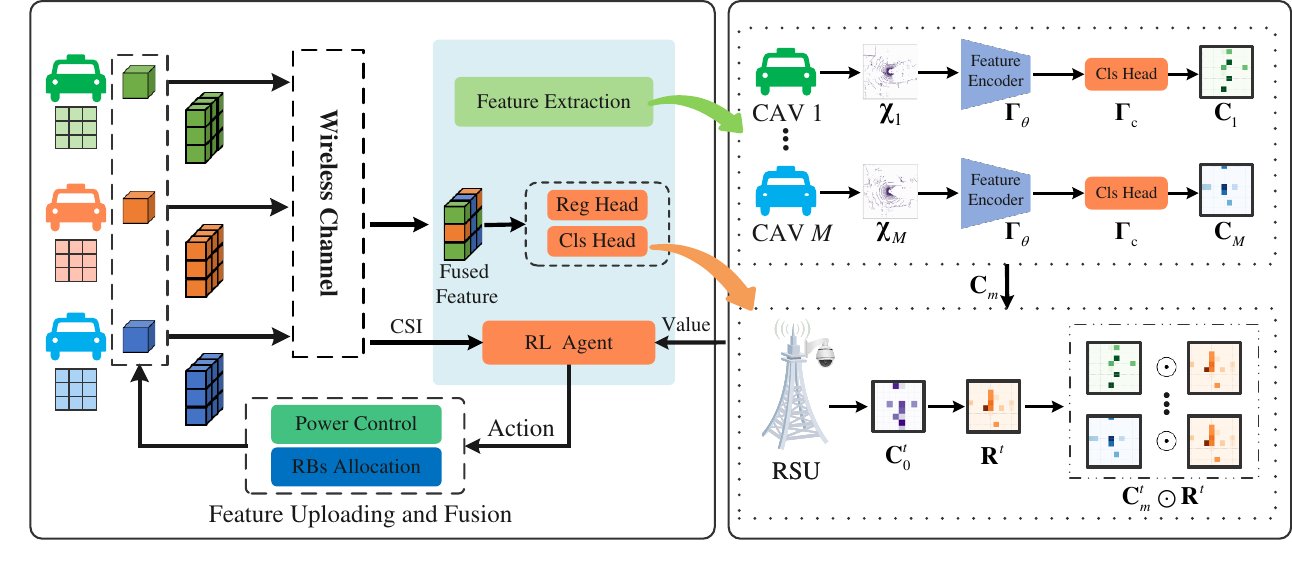}
	\caption{The framework of RACooper.}
	\label{fig_overall}
\end{figure*}

Fig. \ref{fig_overall} shows the proposed RACooper framework, which enables the RSU to interactively learn resource allocation strategies within the environment to improve collaborative perception performance. RACooper involves a two-stage process: initial feature extraction at the start of each period, followed by iterative feature uploading and fusion within each time step. At the beginning of a period, the RSU and all CAVs extract local features to obtain bird's eye view (BEV) representations and generate confidence maps, while CAVs upload their confidence maps to the RSU. Within each time step, the RSU then allocates resources to guide the feature uploads from the CAVs. After the upload, the RSU proceeds with feature fusion and refines the confidence map based on the fused results. This iterative process continues until all CAVs have finished uploading their features or the period ends.

\subsection{Feature Extraction}
In this paper, we take object detection as the downstream task for collaborative perception. Without loss of generality, we adopt PointPillars as the backbone to transform the raw point cloud data into pseudo images \cite{lang2019pointpillars}. To achieve collaborative perception, the RSU periodically broadcasts its extrinsic matrix, which includes spatial coordinates and calibration parameters, allowing CAVs to map their local point clouds to the RSU's unified coordinate system based on rigid body transformation algorithms. Initially, the RSU and CAVs obtain BEV features by passing the collected point cloud data through a feature encoder $\boldsymbol{\Gamma}_{\theta}(\cdot)$. In particular, the BEV features of the $m$th CAV at the beginning of a period is represented as

\begin{equation}
	\mathbf{F}_m = \boldsymbol{\Gamma}_{\theta}(\mathbf{\chi}_{m}) \in \mathbb{R}^{H \times W \times C}, \quad \forall m\in\mathcal{M},
\end{equation}
where $H$ and $W$ represent the height and width, respectively, $C$ is the number of channels, and $\mathbf{\chi}_{m}$ represents the LiDAR point cloud collected by the $m$th CAV. Similarly, the BEV features produced by the RSU are denoted as $\mathbf{F}_0$. 

Based on the BEV features, the CAVs generate local spatial confidence maps by processing those features $\mathbf{F}_m$ through the classification head $\boldsymbol{\Gamma}_{\text{c}}(\cdot)$, which is represented by

\begin{equation}
	\mathbf{C}_m=\boldsymbol{\Gamma}_{\text{c}}(\mathbf{F}_m) \in [0,1]^{H\times W}, \quad \forall m\in\mathcal{M}.
	\label{equ_conf}
\end{equation}
Here the confidence map indicates the likelihood that road participant detects the presence of an object at each coordinate. We assume that each CAV updates its own confidence map solely at the start of every period. In contrast, the RSU, owing to the iterative nature of feature fusion, is required to recompute its confidence map at the beginning of each time step, which is expressed as

\begin{equation}
	\mathbf{C}_0^t =
	\begin{cases}
		\boldsymbol{\Gamma}_{\text{c}}(\mathbf{F}_0), & \text{if } t = 1 \\
		\boldsymbol{\Gamma}_{\text{c}}(\mathbf{F}_\text{fused}^{t-1}), & \text{if } t \neq 1
	\end{cases}
	\quad\forall t \in \mathcal{T},
	\label{conf_r}
\end{equation}
where $\mathcal{T} = \{1,\cdots,T\}$ is the set of step. The process is initialized at $t = 1$ using the local feature $\mathbf{F}_0$. For all subsequent time steps, the confidence map is recursively updated using $\mathbf{F}_\text{fused}^{t-1}$, which represents the features aggregated at the RSU from all participating CAVs during the previous time step.

\subsection{Feature Uploading and Fusion}
In the beginning of $t$th time step, the RSU produces a request map, which is generated based on the local spatial awareness requirements, represented by

\begin{equation}
	\mathbf{R}^{t}=\mathbf{1}-\mathbf{C}_0^{t}\in\mathbb{R}^{H\times W}, \quad \forall t\in\mathcal{T}.
\end{equation}
The request map is negatively correlated with the spatial confidence map. Intuitively, positions with lower confidence scores in $\mathbf{C}_0^{t}$ indicate that the RSU is less certain about the presence of an object at those positions. As a result, these positions need to be complemented by feature from the CAVs to improve perception performance. At the beginning of each period, each CAV transmits its confidence map $\mathbf{C}_m$ to the RSU. Hence, we can compute the perception gain that each CAV provides to the RSU at time step $t$, which is defined as

\begin{equation}
	\mathbf{G}_{m}^{t}=\mathbf{C}_m^{t}\odot\mathbf{R}^{t}, \quad \forall t \in \mathcal{T},  \forall m \in \mathcal{M},
	\label{equ_m}
\end{equation}
where $\odot$ denotes the element-wise multiplication. $\mathbf{C}_m^{t}$ is the confidence map after masking in the previous time step, and we have $\mathbf{C}_m^{t} = \mathbf{C}_m$ if $t = 1$. A higher value at a given position within $\mathbf{G}_{m}^{t}$ indicates a greater complementarity with the RSU. At each time step $t$, the RSU broadcasts $\mathbf{R}^{t}$ before the feature upload, based on which the $m$th CAV computes its corresponding $\mathbf{G}_m^t$ upon reception. Since the data size of $\mathbf{R}^{t}$ is much smaller than that of the perception features, the communication overhead for broadcasting it is considered negligible. Although both the RSU and the CAVs have access to $\mathbf{G}_{m}^{t}$, its roles differ for each. The RSU utilizes it to evaluate the perception gain from each CAV, which informs resource allocation decisions. For the CAVs, it highlights which features are most beneficial to the RSU's perception, thus guiding the feature selection process for uploading. Accordingly, the $m$th CAV selects features to upload from its local BEV features based on the link capacity and the $\mathbf{G}_{m}^{t}$, denoted as

\begin{equation}
	\mathbf{M}_m^t = \mathcal{F}_\text{top}( \mathbf{G}_{m}^{t} \big| \lfloor B_m^t \rfloor)\in\{0, 1\}^{H\times W}, \quad \forall t \in \mathcal{T}, \forall m \in \mathcal{M},
	\label{equ_mask}
\end{equation}
where $\mathbf{M}_m^t(h,w)  = 1$ represents that $\mathbf{F}_m(h,w)$ is chosen for transmission and $\mathbf{M}_m^t(h,w)  = 0$ otherwise. $B_m(t)$ is the maximum number of features that the $m$th V2I link can upload at the $t$th time step, which is influenced by the resource allocation decision and will be discussed in Section III. The function $\mathcal{F}_\text{top}( \cdot \big| \lfloor B_m^t \rfloor)$ identifies the top $\lfloor B_m(t) \rfloor$ points by value, assigning $1$ to their corresponding locations on $\mathbf{M}_m^t$ and $0$ elsewhere. Here, $\lfloor x \rfloor = \max \left\{ n \in \mathbb{Z} \mid n \leq x \right\}$ denotes the floor function, which rounds $x$ down to the nearest integer. At the end of time step $t$, 
both the RSU and each CAV update their respective local confidence maps in (\ref{equ_m}) by masking out transmitted features to avoid redundant transmission, which is represented by
\begin{equation}
	\mathbf{C}_m^{t+1} = \mathbf{C}_m \odot (\mathbf{1} - \sum_{\tau=1}^{t} \mathbf{M}_m^{\tau}), \quad \forall t \in \mathcal{T}, \forall m \in \mathcal{M}.
	\label{c_mt}
\end{equation}

After several rounds of feature transmission, the accumulated uploaded features $	\mathbf{A}_{m}^{t} \in \mathbb{R}^{H\times W \times C}$ of the $m$th CAV received by the RSU at step $t$ is given by

\begin{equation}
	\mathbf{A}_{m}^{t} = \left( \sum_{\tau=1}^{t} \mathbf{M}_{m}^{\tau} \right) \odot \mathbf{F}_m, \quad \forall t \in \mathcal{T}, \forall m \in \mathcal{M}.
\end{equation}
Subsequently, the RSU merges all features from CAVs with its own BEV features to produce the $t$th step fused features in (\ref{conf_r}), which is represented by

\begin{equation}
	\mathbf{F}_\text{fused}^{t}=\boldsymbol{\Gamma}_{\varphi}(\{\mathbf{A}_{m}^{t}\}_{m \in \mathcal{M}}, \mathbf{F}_0), \quad \forall t \in \mathcal{T},
\end{equation}
where $\boldsymbol{\Gamma}_{\varphi}(\cdot)$ represents the fusion network. Upon experiencing all the steps within the period, the RSU acquires the final collaborative features. These features are then fed into the classification head $\boldsymbol{\Gamma}_{\text{c}}(\cdot)$ and regression head $\boldsymbol{\Gamma}_{\text{r}}(\cdot)$, where $\boldsymbol{\Gamma}_{\text{r}}(\cdot)$ predicts the three-dimensional position, size, and yaw angle of the bounding box, with $\boldsymbol{\Gamma}_{\text{c}}(\cdot)$ produces a confidence score for each bounding box to indicate the likelihood of containing an object for each bounding box. Finally, post-processing is applied to both the regression outputs and the confidence scores to derive the final detection results $\hat{y}$.

\section{Resource allocation in RACooper}

\subsection{Transmission Model}
It is essential to recognize that communication interference is inevitable, but through effective resource allocation strategies, the number of perception features received by the RSU can be significantly increased, thereby enhancing the accuracy of collaborative perception. By employing orthogonal frequency division multiplexing (OFDM) technology, a frequency selective wireless channel can be converted into multiple flat fading channels over the subcarriers. The total system bandwidth can be divided into a set of orthogonal resource blocks (RBs), denoted by $\mathcal{K}=\{1,2,\cdots,K\}$. These RBs are shared among all CAVs. It is posited that the channel fading remains relatively uniform within each RB, and experiences noticeable change across different RBs. For the $m$th V2I link at the $t_s$th sub-time step of the time step $t$, the channel power gain on the $k$th RB is represented as follows

\begin{equation}
	g_m^k(t, t_s) = \alpha_m |h_m^k(t, t_s)|^2,
\end{equation}
where $\alpha_m$ represents frequency-independent large-scale fading, encompassing both path loss and shadowing, while \( h_m^k(t, t_s) \) represents  frequency dependent small-scale fading. The utilization of these RBs is represented by \( \eta_m^k(t) \in \{0,1\} \) for each RB $k\in\mathcal{K}$ and and each V2I link $m\in\mathcal{M}$, where $\eta_m^k(t) = 1$ if RB $k$ is allocated to the $m$th V2I link, and $\eta_m^k(t) = 0$ otherwise. Each V2I link is associated with at most one RB, i.e., $\sum_{k\in\mathcal{K}}\eta_{m}^k(t) \leq 1 $, for all $m\in \mathcal{M}$, but each RB may be assigned to multiple links. Hence, when the $m$th V2I link is assigned the $k$th RB at the $t_s$th sub-time step of step $t$, the interference power to the $m$th link can be represented by

\begin{equation}
	I_m^k(t, t_s) = \sum\limits_{j\in \mathcal{M},j\neq m} P_j(t) \cdot g_{j}^k(t, t_s) \cdot \eta_j^k(t), \quad \forall t \in \mathcal{T},
\end{equation}
where $P_j(t)$ is the transmit power of the $j$th CAV at time step $t$. Consequently, the total transmission rate from the $m$th CAV to the RSU across all RBs in the set $\mathcal{K}$ at the sub-time step $t_s$ of time step $t$ is expressed as

\begin{equation} \label{eq:total_rate} R_m(t, t_s) = \sum_{k \in \mathcal{K}} W_B \cdot \eta_m^k(t) \cdot \log_2\left(1 + \frac{P_m(t) \cdot g_{m}^k(t, t_s)}{I_m^k(t, t_s) + \sigma^2}\right), 
\end{equation}
where $W_B$ is the bandwidth of each RB,  and \( \sigma^2  = N_0W_B\) is the noise power with $N_0$ being the power spectral density of the additive white Gaussian noise. The number of features uploaded by the $m$th CAV at the sub-time step $t_s$ of time step $t$ is represented by

\begin{equation}
	B_m(t,t_s) = \frac{R_m(t, t_s) \times \Delta t_s}{C \times Q}, 
	\label{equ_Bmt}
\end{equation}
with $Q$ is the quantization precision of each BEV and $\Delta t_s$ is the duration of each sub-time
step, i.e., the channel coherence time. Finally, the number of uploaded features by the $m$th CAVs in (\ref{equ_mask}) in each time step $t$ is given by
\begin{equation}
	B_m(t) = \sum\limits_{t_s = 1}^{T_s} {B_m(t,t_s)}.
	\label{totalB}
\end{equation}

\subsection{Problem Statement}

The core challenge in RACooper is to design an efficient resource allocation strategy tailored for collaborative perception. The allocation of limited communication resources must maximize perception performance. To this end, the optimization objective of RACooper is to maximize the global perception accuracy of the RSU in collaborative scenarios subject to constraints on communication resources. For generality, we use average precision (AP) as the performance metric. Specifically, the optimization is designed to allocate resources efficiently based on CSI and collaborators' contribution to overall perception accuracy. The optimization variables include the spectrum allocation $\eta_{m}^k(t)$ and transmit power $P_m(t)$ for $k\in \mathcal{K}$ and $m\in\mathcal{M}$. Hence, we present the optimization problem for RACooper as

\begin{align}
	&\max_{\{ \eta_m^k(t), P_m(t)\}_{k\in\mathcal{K}, m\in\mathcal{M}, t\in\mathcal{T}}} \text{AP}\big(\hat{y}, y_\text{true}\big) \label{equ_obr} \\
	\text{s.t.} \quad &\sum_{k\in\mathcal{K}}\eta_{m}^k(t) \leq1, \quad \forall m\in\mathcal{M},\forall t \in \mathcal{T},  \tag{\ref{equ_obr}a} \label{Problema} \\
	&0\leq P_m(t)\leq P_\text{max}, \quad \forall m\in\mathcal{M},  \forall t \in \mathcal{T}, \tag{\ref{equ_obr}b} \label{Problemc} \\
	&\sum_{h, w}\mathbf{M}_{m}^{t}(h, w)\leq  {\lfloor B_m(t) \rfloor}, \quad \forall m\in\mathcal{M},\forall t \in \mathcal{T}, \tag{\ref{equ_obr}c} \label{Problemb}
\end{align}
where $h$ and $w$ represent the horizontal and vertical coordinates, respectively, and $y_\text{true}$ represents the ground truth. Additionally, a transmit power budget, $P_\text{max}$, constraint is imposed on each CAV.
 
Due to the complexity of calculating AP and the real-time requirements of the collaborative perception system, directly addressing the problem (\ref{equ_obr}) in RACooper is challenging. Therefore, we adopt a more tractable approach by minimizing a surrogate loss function instead. For object detection tasks, the loss function can be represented by

\begin{equation}
	{L}_\text{det} =\lambda_\text{cls}{L}_\text{cls}+\lambda_\text{loc}{L}_\text{loc}+\lambda_\text{dir}{L}_\text{dir},
	\label{equ_loss}
\end{equation}
where \( {L}_\text{cls} \) represents the the classification loss, \( {L}_\text{loc} \) represents the localization loss, and \( {L}_\text{dir} \) represents the orientation loss. And the terms \( \lambda_\text{cls} \), \( \lambda_\text{loc} \),  and \( \lambda_\text{dir} \) are the weights for the respective loss components. The goal of \( {L}_\text{cls} \) is to improve the accuracy of classification confidence, while the objectives of \( {L}_\text{loc} \) and \( {L}_\text{dir} \) are to enhance the target intersection over union (IoU) quality. Therefore, minimizing the function defined in (\ref{equ_loss}) helps to achieve more accurate classification, localization, and orientation, which further enhances the AP performance. To this end, the final optimization problem in RACooper can be reformulated as

\begin{equation}
	\begin{aligned}
	\min_{\{ \eta_m^k(t), P_m(t)\}_{k\in\mathcal{K}, m\in\mathcal{M}, t\in\mathcal{T}}} {L}_\text{det} \label{equ_ob} \\
	\text{s.t.} \quad  (\ref{Problema}), (\ref{Problemc}), (\ref{Problemb}).
	\end{aligned}
\end{equation}

It is noteworthy that this optimization problem differs significantly from traditional resource allocation problems. Conventional resource allocation aims to find a scheme that optimizes specific communication metrics, such as sum rate, bit error rate, or latency, based on the current channel state. In contrast, this formulation is designed to achieve the different goal of maximizing long term perception performance. This may involve prioritizing the transmission of data that is more valuable for collaborative perception, even if such a strategy does not yield the highest instantaneous communication rates. However, solving the optimization problem in (\ref{equ_ob}) with conventional methods is highly challenging. A primary challenge is that the relationship between a resource allocation decision and its ultimate impact on the long term perception performance is too complex to be captured by a tractable analytical model. Fortunately, deep reinforcement learning (DRL) has been shown powerful in learning policies to improve long-term performance through interacting with the unknown environment. Therefore, we employ a DRL-based method to address the resource allocation problem \cite{zhao2023adaptive, an2024channel}, and the details are discussed in the next section.

\section{Hierarchical RL-based Resource Allocation Design}
In RACooper, the RSU functions as the agent, learning optimal resource allocation policy through its interaction with the environment. To manage the complexity of the joint decision-making process, we design the agent using a hierarchical RL architecture. This approach breaks down the overall resource allocation task into two sub-tasks: spectrum allocation and power control, which are addressed in sequence. Two separate DRL models are used to learn the policies for each sub-problem in a hierarchical manner. At each step $t$, the agent first determines spectrum allocation by assigning RBs to the V2I links, and then selects the appropriate transmit power for each V2I link to minimize the object detection loss in the RSU. This decomposition simplifies the learning challenge at each level of the hierarchy, enabling a more structured exploration of the policy space. The key components of the hierarchical RL-based resource allocation design are outlined below in detail.

\subsection{Action Space}
Following the hierarchical RL framework, the agent's action is a composite of decisions from both hierarchical levels. The joint action at step $t$ can be represented as

\begin{equation}
	a(t) = \{\eta_m^k(t), P_m(t)\}_ {m \in \mathcal{M}, k \in \mathcal{K}}.
\end{equation}
For the ease of policy learning, we limit the power control options to three levels, i.e., \([23, 10.5, -100]\) dBm, where $-100$ dBm corresponds to zero power.

\subsection{State Space}

Conceivably, the resource allocation for enhancing collaborative perception performance in RACooper is influenced by two factors. The first is the CSI of the V2I links. Superior channel conditions translate to higher achievable data rates, which permit the transmission of more features from a CAV to the RSU. This increased data volume has the potential to improve the accuracy of the final collaborative perception. The perceptual gains of the features is equally important. Features from different CAVs contribute differently to the RSU, with some features being critical and others redundant. Therefore, features that are more complementary to those acquired at the RSU yield higher perceptual gains.

As mentioned above, the values in $\mathbf{G}_{m}^{t}$ complement the confidence map at the RSU, capturing the contribution level of each CAV with RSU. Therefore, we define the feature value of CAV \(m\) for the RSU at each step $t$ as

\begin{equation}\label{equ_v}
	v_{m}^{t}= \sum_{h,w} \mathbf{G}_m^t(h, w).
\end{equation}
Inspired by the previous perspective, the input states for the RB allocation DRL model and power control DRL model at time step $t$ are given by
\begin{equation}
	\begin{aligned}
		s_{\eta}(t)&=\{\{\alpha_m, h_m^{t}\},v_m^{t}\}_{m\in\mathcal{M}},\\
		s_p(t)&=\{\{\alpha_m, h_m^{t}\},v_m^{t}, \eta_m^k(t)\}_{m\in\mathcal{M}, k \in \mathcal{K}}.
	\end{aligned}
	\label{equ_s}
\end{equation}

Following time step $t$, the agent experiences a state transition in the environment, which includes the update of fast fading, the generation of new confidence map by (\ref{conf_r}), and the masking of features transmitted during this time step.

\subsection{Reward Design}
Our objective is to minimize the detection loss ${L}_\text{det}$ in RACooper by guiding the agent with a reward signal. One direct approach to formulating the reward is to base it on the change in this loss. However, in collaborative perception systems, the loss values of neural networks can be highly non-smooth and noisy, resulting in significant fluctuations. This makes it challenging for RL agents to learn a stable and generalizable policy. To address this challenge, we impart our domain knowledge by complementing the change in loss with the communication rates in the reward function. As mentioned in the state design, higher communication rates ensure more feature transmission, enhancing the RSU’s detection performance. Accordingly, the reward function is formulated as
\begin{equation}\label{equ_r}
	R_t=\lambda_\text{rate}\sum_{m=1}^M R_m(t)-\lambda_\text{det}\Delta{L}_\text{det},
\end{equation}
where \(\Delta{L}_\text{det}\) represents the change in loss after each decision, with \(\lambda_\text{rate}\) and \(\lambda_\text{det}\) are weights introduced to balance the communication rates and the loss function components. By incorporating both communication efficiency and detection loss reduction, the reward function provides a more informative signal for policy learning. Therefore, RACooper is fundamentally a communication optimization framework driven by perception tasks, where the agent strikes a balance between communication rates and perception performance through ongoing interaction with the environment.

\subsection{Hierarchical RL-Based Resource Allocation Algorithm}

Each episode in the RL formulation is set to the sampling or collaborative perception period, while the duration of each step corresponds to the interval for each resource allocation adjustment. The transmission of features and variations in small-scale fading drive transitions in the environment's state. The proposed hierarchical RL framework utilizes proximal policy optimization (PPO) as its policy optimization method for both spectrum allocation and power control \cite{schulman2017proximal}.

\begin{figure}
	\centering
	\includegraphics[width=\linewidth]{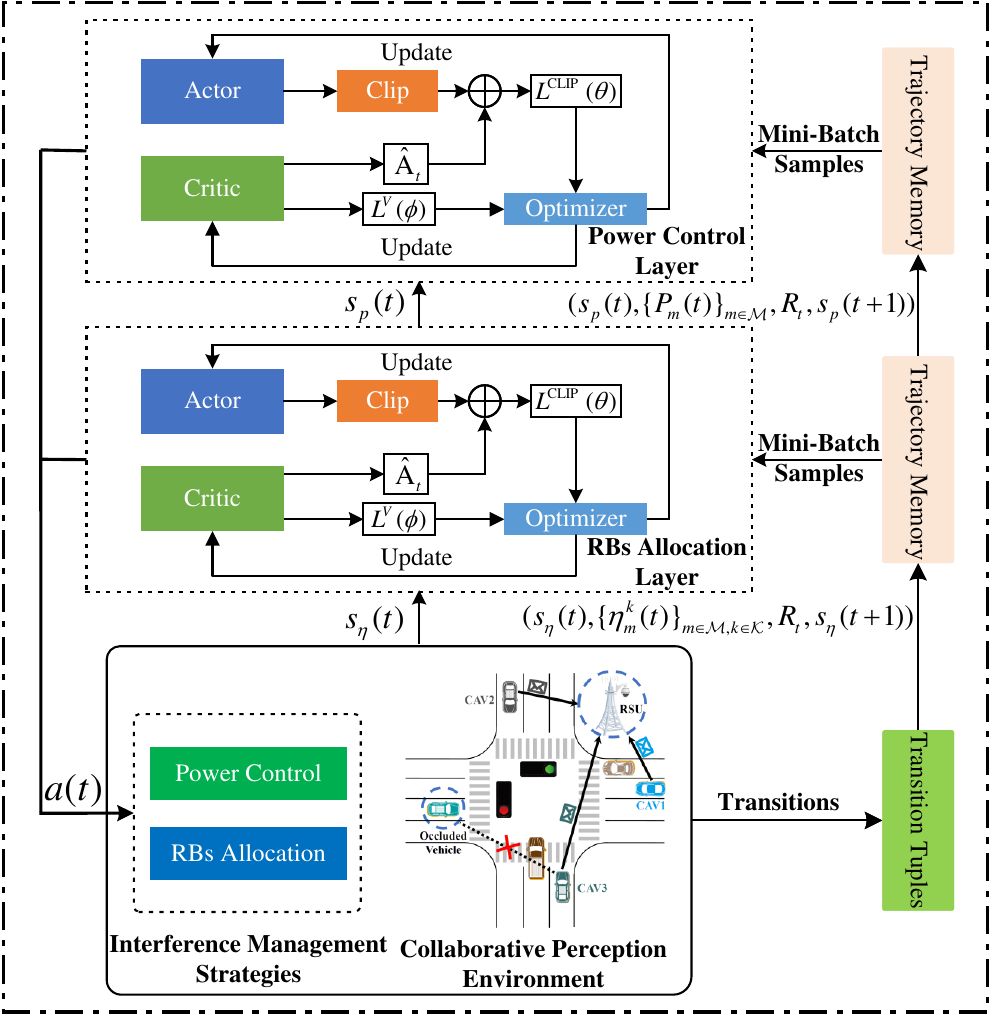}
	\caption{The agent-environment interaction in RACooper formulation of the investigated resource allocation in collaborative perception system.}
	\label{fig_rlp}
\end{figure}

Our hierarchical PPO (HPPO) consists of two parts, training and implementation. As shown in Fig. \ref{fig_rlp}, the HPPO training process begins by initializing environment variables (i.e., channel fading model and collaborative perception scenarios) and the hierarchical RL model. At each step $t$, the PPO model in the RB allocation layer observes the state \(s_{\eta}(t)\) from the environment and executes a spectrum allocation action 
\(\{\eta_m^k(t)\}_{m \in \mathcal{M}, k \in \mathcal{K}}\) from the action space. The chosen spectrum allocation action, along with the state \(s_{\eta}(t)\), are concatenated to form a state \(s_p(t)\), for the PPO model in the power control layer. In this layer, the model selects a power allocation action \(\{P_m(t)\}_{m \in \mathcal{M}}\) from its own action space. Once both layers have executed their respective actions, the environment provides a reward \(R_{t+1}\) based on the joint action $a_t$ and transits to new states \(s_{\eta}(t+1)\) and \(s_p(t+1)\), with \(s_p(t+1)\) determined only after the RB allocation action is made, following the update of \(s_{\eta}(t+1)\). Due to environment changes caused by channel variations and the actions taken by the agent, the agent collects and stores transition tuples, $\{s_{\eta}(t),\{\eta_m^k(t)\}_{m \in \mathcal{M}, k \in \mathcal{K}},R_{t},s_{\eta}(t+1)\} $ and $\{s_p(t),\{P_m(t)\}_{m \in \mathcal{M}},R_{t},s_p(t+1)\}$, in the trajectory memory of each layer. The environment’s feedback rewards are then buffered to refine the policy, while the agent continues executing actions to adapt the environment.

Once a sufficient number of trajectories are collected, the model begins policy optimization. HPPO uses this experience data to compute advantage estimates, which are subsequently used to update the policy network. For each policy update, a batch of these trajectories is sampled to update the network parameters. To address the high variance and instability inherent in the sampling process, PPO adopts a clipped surrogate objective function that constrains the policy update step size, thereby ensuring stable learning. This objective is formulated as

\begin{equation}\label{equ_p}
	L^{\mathrm{CLIP}}(\theta)=\hat{\mathbb{E}}_t[\min(r_t(\theta)\hat{A}_t,\mathrm{clip}(r_t(\theta),1-\epsilon,1+\epsilon)\hat{A}_t)],
\end{equation}
where \(r_t(\theta) = \frac{\pi_\theta(a_t|s_t)}{\pi_{\theta_{\mathrm{old}}}(a_t|s_t)}\) denotes the ratio of probabilities between the new and old policies. The term \(\hat{A}_t = Q(s_t, a_t) - V(s_t)\) represents the advantage function, quantifies the relative merit of an action $a_t$ compared to the baseline value of the state. The state value function $V(s_t)$ is approximated by the critic network, whose parameters $\phi$ are updated by minimizing the mean squared error loss

\begin{equation}
	\label{l_critic}
	\setlength\abovedisplayskip{10pt}
	\setlength\belowdisplayskip{6pt}
	{L^V}(\phi) =\mathbb{E}_t\left[\left(V_\phi\left(s_t\right)-R_t\right)^2\right].
\end{equation}

We maintain consistency in parameters for different layers' PPO model, except for the input and output layers. Although the full action $a_t$ is jointly determined by two actors, each actor updates independently at each update phase according to its own update rules, as per (\ref{equ_p}). The training procedure is summarized in Algorithm 1.

\begin{algorithm}[t]
	\caption{Hierarchical DRL-Based Resource Allocation Algorithm for V2I Collaborative Perception}
	\begin{algorithmic}[1]
		\State Initialize HPPO network for the agent;
		\State Initialize the environment variables;
		\State Initialize the backbone network with pre-trained weights and freeze its parameters;
		\For{each period}
		\State Load a new episode from dataset;
		\State The RSU and all CAVs generate the spatial 
		\State confidence map; 
		\State Update locations and large-scale fading $\alpha$;
		\For{each step $t$}
		\State Obtain the observation $s_{\eta}(t)$ and 
		\State execute action $\{\eta_m^k(t)\}_{m \in \mathcal{M}, k \in \mathcal{K}}$;
		\State Obtain the observation $s_p(t)$ and 
		\State execute action $\{P_m(t)\}_{m\in\mathcal{M}}$;
		\For{each sub-time step $t_s$ in step $t$}
		\State Update channel small-scale fading;
		\EndFor
		\State CAVs transmit features to RSU;
		\State Perform feature fusion at RSU;
		\State Calculate the reward $R_{t}$;
		\State Obtain new state $s_{\eta}(t+1)$ and $s_p(t+1)$;
		\State Collect trajectories 
		\State $\{s_{\eta}(t),\{\eta_m^k(t)\}_{m \in \mathcal{M}, k \in \mathcal{K}},R_{t},s_{\eta}(t+1)\}$ and
		\State $\{s_p(t),\{P_m(t)\}_{m \in \mathcal{M}},R_{t},s_p(t+1)$;
		\EndFor
		\If{every 10 episodes}
		\State Update networks in spectrum allocation layer;
		\State Update networks in power control layer;
		\EndIf
		\EndFor
		\label{Alg_1}
	\end{algorithmic}
\end{algorithm}

The object detection backbone uses the architecture from \cite{li2022v2x}, which incorporates $\boldsymbol{\Gamma}_{\theta}(\cdot)$, $\boldsymbol{\Gamma}_{\text{c}}(\cdot)$, and $\boldsymbol{\Gamma}_\text{r}(\cdot)$ from PointPillars, along with the fusion module $\boldsymbol{\Gamma}_{\varphi}(\cdot)$. For our experiments, we employ a pre-trained network and keep all its parameters frozen. As a result, both the training and implementation phases are simplified, focusing solely on the optimization of the HPPO. During the implementation phase, the agent receives the observation \( s_{\eta}(t) \) and \( s_p(t) \) at each step $t$. It then selects the actions corresponding to the maximum output value from the the spectrum allocation layer and the power control layer, respectively.

\section{Simulation Result}
In this section, we provide simulation results to demonstrate the effectiveness of RACooper. The experiments are conducted using the V2X-Sim dataset \cite{li2022v2x}, which is utilized for both training and testing RACooper. The performance is evaluated based on two metrics: AP at an IoU threshold of 0.5 (AP@0.5) and AP at an IoU threshold of 0.7 (AP@0.7). Note that we selectively utilize scenarios from the V2X-Sim dataset because the maximum number of CAVs in some scenarios does not align with our assumptions. Specifically, we concentrate on scenarios featuring intersections with one deployed RSU and a precise configuration of four collaborating CAVs. The sensory streams within the V2X-Sim dataset are recorded at a frequency of 5 Hz, hence we set a collaborative perception period to 200 ms, which also corresponds to one episode in the RL process.

\begin{table}[t]
	
	\begin{center}
		\normalsize
		\renewcommand{\arraystretch}{1}
		\caption{Simulation Parameters}
		\label{tab1}
		\begin{tabular}{l | p{3cm}<{\centering} }
			\hline
			
			\textbf{Parameters} & \textbf{Values} \\
			\hline
			
			Number of V2I links $M$& 4 \\
			
			Number of subcarriers RBs $K$& 2 \\ 
			
			Carrier frequency & 5.9 GHz\\
			
			Bandwidth & $2.5 \sim 3.5$ MHz  \\
			
			RSU antenna height & 25 m\\
			
			RSU antenna gain & 8 dBi\\
			
			RSU receiver noise figure & 5 dB\\
			
			Vehicle antenna gain & 3 dBi \\
			
			Vehicle receiver noise figure & 9 dB \\

			Absolute vehicle speed $v$ & 0 km/h $\sim$ 25 km/h \\ 
			
			V2I transmit power & [23,10.5,-100] dBm \\
			
			Collaborative perception period & 200 ms
			\\
			
			Step duration & 5 ms\\
			
			Small-scale fading update interval & 1 ms
			\\
			
			Noise power $\sigma^2$ & -114 dBm \\
			\hline
		\end{tabular}
	\end{center}
\end{table}

\begin{table}[t]
	\begin{center}
		
		\normalsize
		\renewcommand{\arraystretch}{1}
		\caption{Parameters for Model Training}
		
		\begin{tabular}{l | p{3cm}<{\centering} }
			\hline
			
			\textbf{Parameters} & \textbf{Values} \\
			\hline
			
			Training frames & 2578 \\
			
			Testing frames & 300 \\
			
			Validation frames & 15 \\
			
			Training episodes &  20,000 \\ 
			
			Validation episodes & 3,000 \\
			
			Clipping probability ratio $\epsilon$ & 0.2 \\
			
			Weights $\lambda_\text{det}$ and $\lambda_\text{rate}$ &  0.025 and 20 \\
			
			\hline
		\end{tabular}
	\end{center}
\end{table}

\subsection{Simulation Settings}

To simulate the communication environment, we utilize the raw data from V2X-Sim dataset to construct a V2X communication environment based on 3GPP TR 38.885 \cite{3gpp2018nr} to support simulation experiments. As presented in Table I, we set the update period for large-scale channel fading to 200 ms (i.e., a period), while small-scale channel fading is updated every 1 ms (i.e., a sub-time step). The resource allocation decision interval is configured to 5 ms, resulting in 40 decision intervals per episode. Additionally, the C-V2X system uses 2 RBs, with bandwidth values varying from 2.5 MHz to 3.5 MHz in 0.1 MHz increments. To evaluate the model's robustness and generalization capabilities, we train it using a 3 MHz bandwidth. The remaining bandwidth values within this range are then used solely for the testing phase.

The hyperparameters used for training are detailed in Table II. For our experiments, the V2X-Sim dataset is partitioned into three mutually exclusive sets: a training set of 2,578 frames, a validation set of 15 frames, and a test set of 300 frames. The agent is trained for a total of 20,000 episodes on the training set. To monitor convergence, we evaluate the model's performance on the validation set every 100 episodes. After training is complete, the final model is rigorously evaluated on the unseen test set to report its performance. The PPO model utilizes a three-layered fully-connected multi-layer perceptron network with neuron counts of 500, 250, and 125 in each consecutive layer. Each linear layer is succeeded by a rectified linear unit non-linear activation function. The learning rates for both the actors and critics are configured to \(1 \times 10^{-4}\) and \(3 \times 10^{-4}\), respectively. Furthermore, the hyperparameter clipping probability ratio \(\epsilon\) is set to 0.2. The weights of communication rates and perception performance \(\lambda_\text{rate}\) and \(\lambda_\text{det}\) in (\ref{equ_r}) are 0.025 and 20, respectively. Unless stated otherwise, the simulation parameters for the following experiments default to the values specified in Tables I and II.

In the following, we compare the proposed algorithm with three baselines:

\begin{enumerate}
	\item \textbf{Random}: Each V2I link randomly selects an RB and upload features to the RSU at a random power level.
	
	\item \textbf{Max Rate}: Exhaustively searching iteration over RBs and transmission power, the V2I links with the best channel state and highest 
	sum rate are selected to transmit features.
	
	\item \textbf{Max Features}: In each time, the two CAVs with the most remaining features are selected to access different RB and transmit features to the RSU at max power level.
\end{enumerate}

\subsection{Convergence and Effectiveness}

\begin{figure}[t]
	\centering
	\includegraphics[width=0.9\linewidth]{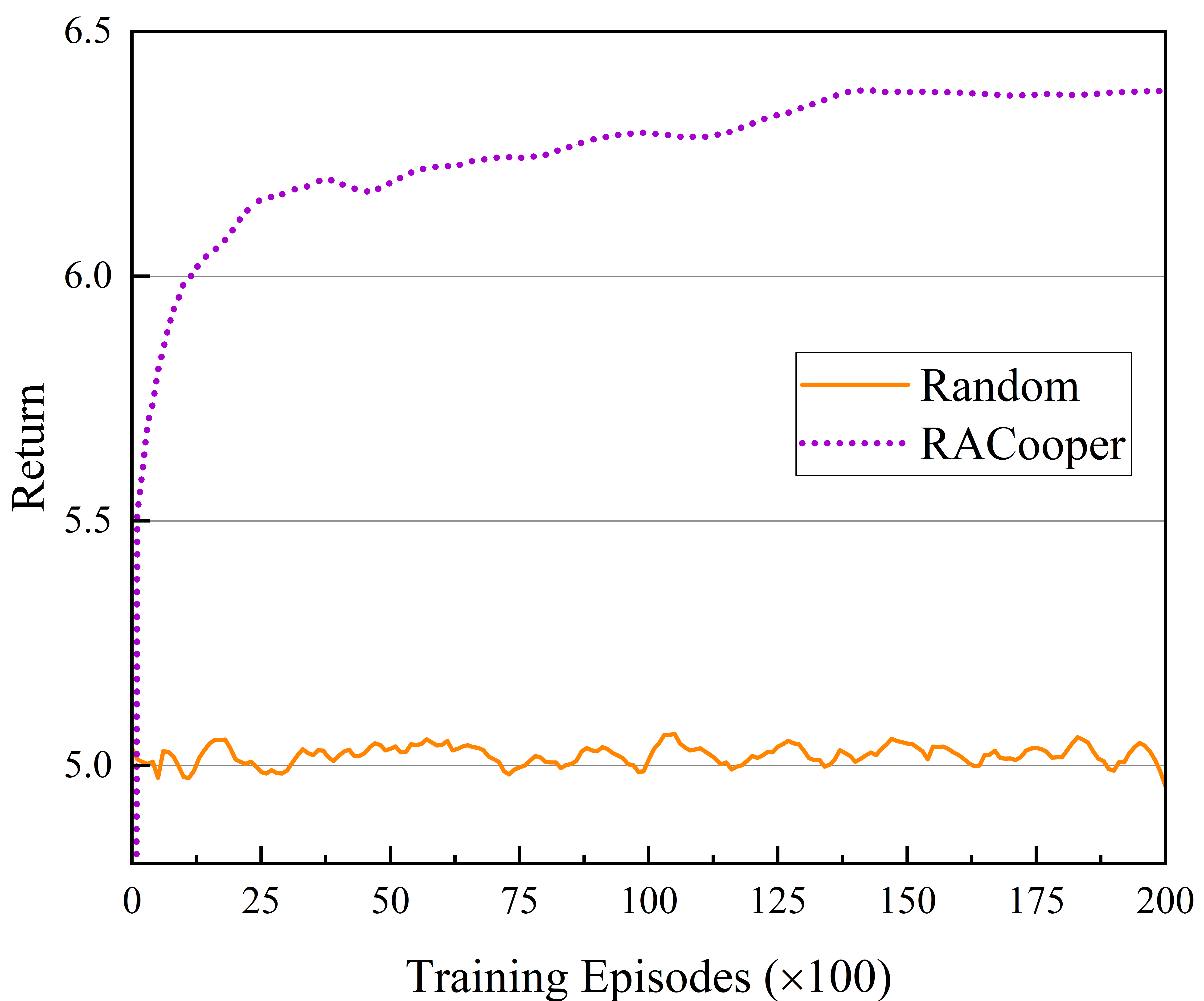}
	\caption{Return on the validation set with increasing iterations.}
	\label{fig_re}
\end{figure}

Firstly, we examine the convergence behavior of the proposed HPPO. Due to the complexity of perception and communication environments in collaborative perception scenarios, the RL algorithm’s returns vary across different episodes. To plot the return curve, we calculate the average return using 15 frames from the validation set every 100 training episodes. Fig. \ref{fig_re} illustrates the return as a function of the number of training iterations. As shown, the HPPO outperforms the Random baseline, effectively learning a suitable policy through interaction. The return demonstrates a steady upward trend and reaches convergence, highlighting the HPPO's favorable convergence characteristics.

\begin{figure}[t]
	\centering
	\includegraphics[width=0.85\linewidth]{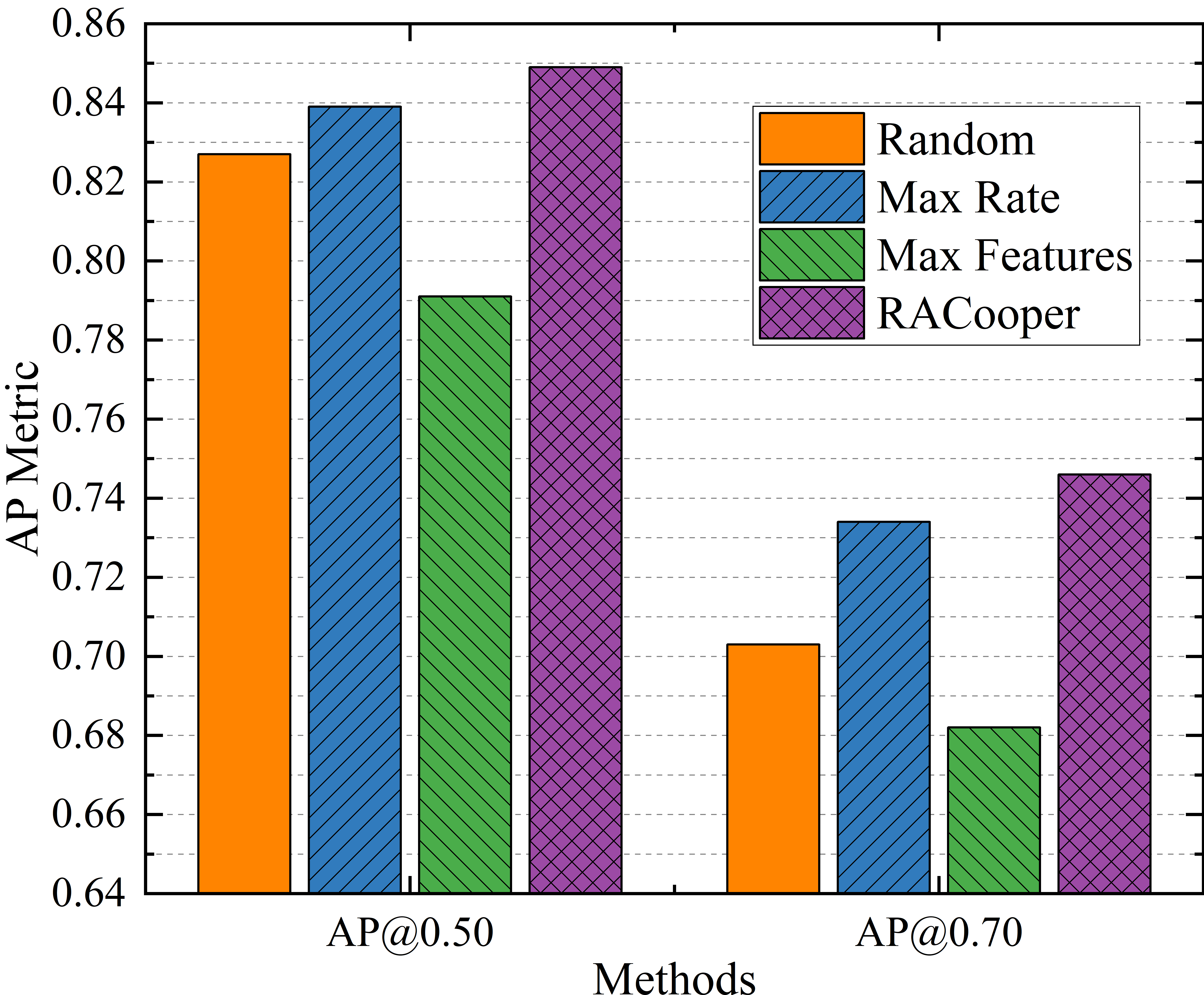}
	\caption{AP performance of different methods under 3 MHz bandwidth.}
	\label{fig_ap3}
\end{figure}

\begin{figure}[tb]
	\centering
	\subfigure[AP@0.50 metric.]{
		\includegraphics[width=0.85\linewidth]{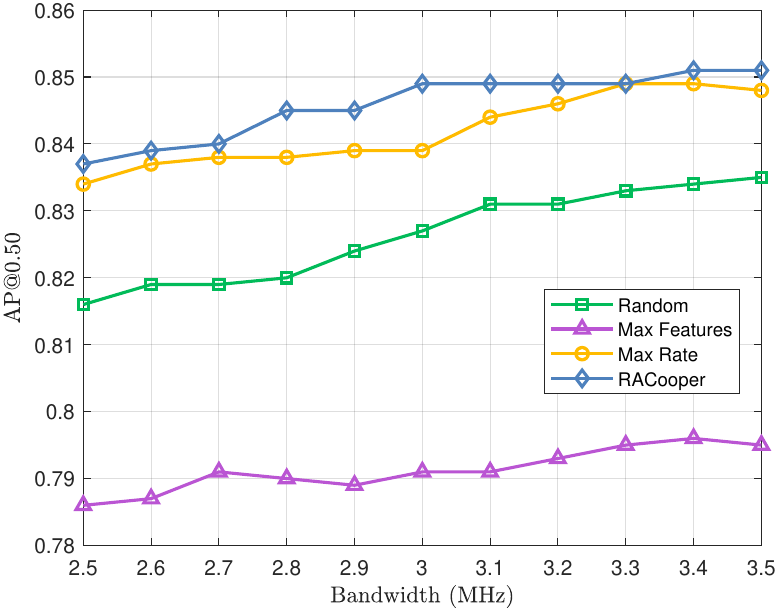}
		\label{fig_0.5} 
	}
	\subfigure[AP@0.70 metric.]{
		\includegraphics[width=0.85\linewidth]{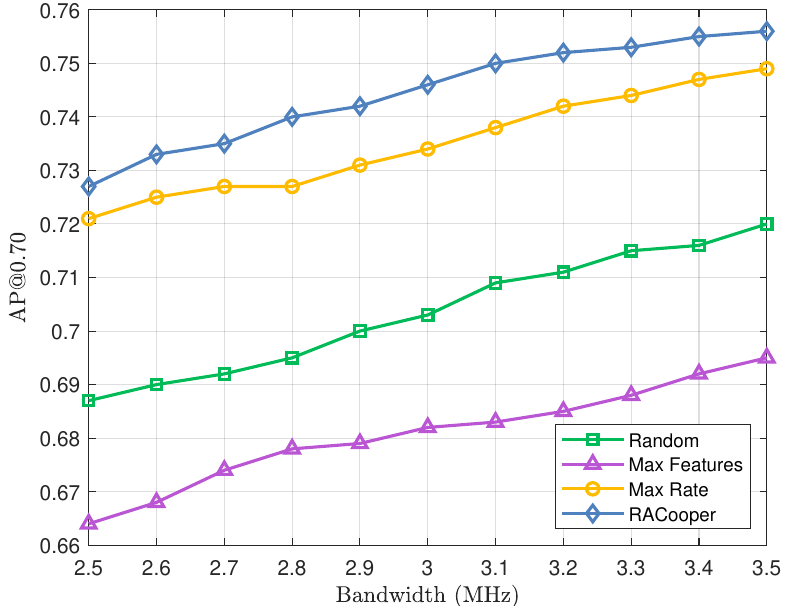}
		\label{fig_0.7} 
	}
	\DeclareGraphicsExtensions.
	\caption{The comparison of various resource allocation methods in different AP metrics.}
	\label{fig_ap}
\end{figure}

\begin{figure}[tb]
	\centering
	\subfigure[Detection loss.]{
		\includegraphics[width=0.85\linewidth]{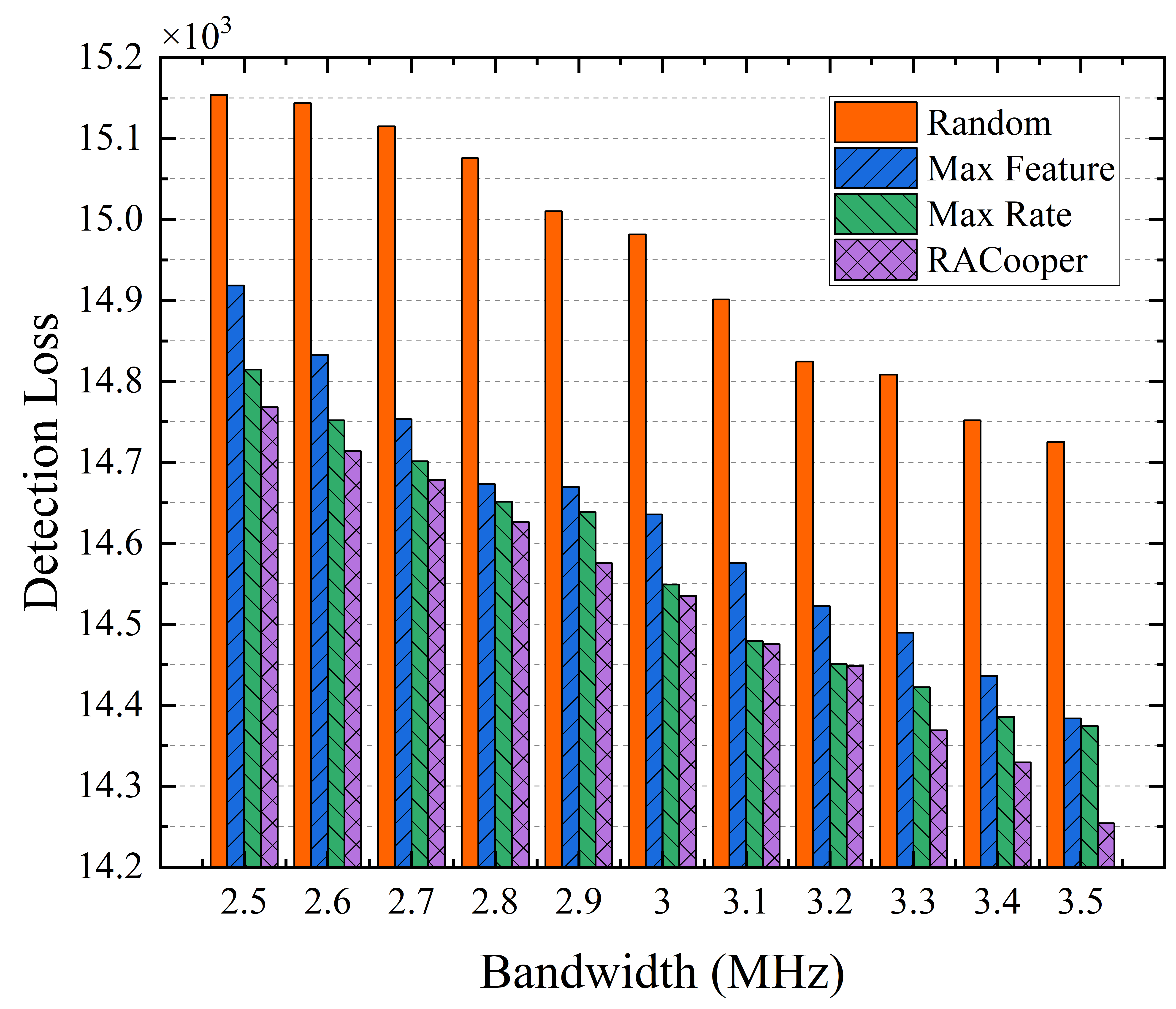}
		\label{fig_det_loss} 
	}
	\subfigure[Classification loss.]{
		\includegraphics[width=0.85\linewidth]{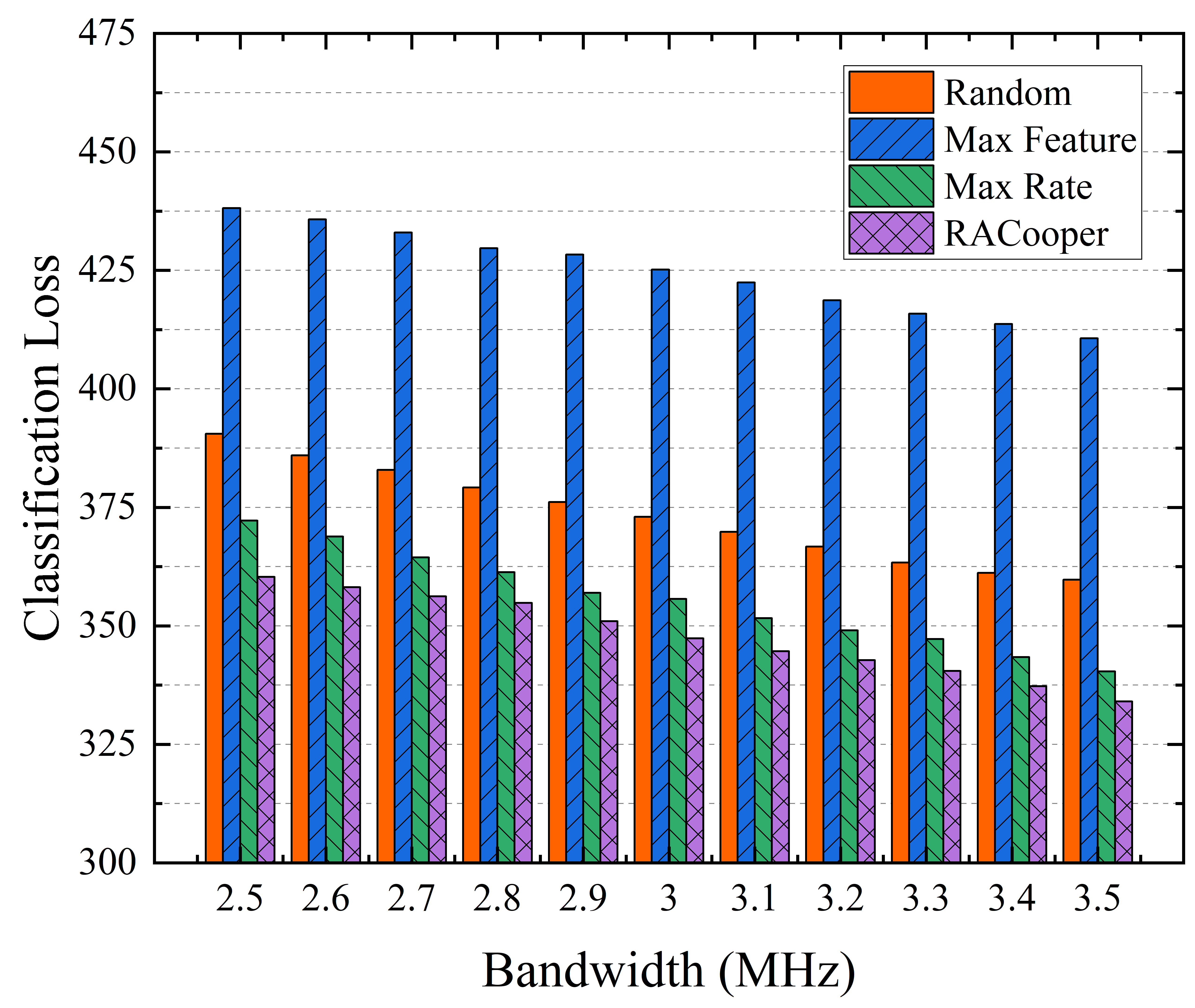}
		\label{fig_cls_loss} 
	}
	\DeclareGraphicsExtensions.
	\caption{Detection and classification loss of different methods under various training bandwidth.}
	\label{fig_loss}
\end{figure}

To evaluate the effectiveness of RACooper, we compare its perception performance with above baseline methods. RACooper was trained on a 3 MHz bandwidth and tested across the 2.5 $\sim$ 3.5 MHz range with 0.1 MHz step increments. Fig. \ref{fig_ap3} show the AP performance of RACooper and other resource allocation methods under 3 MHz bandwidth. As shown in Fig. \ref{fig_ap3}, RACooper achieves performance improvements of 1.2\% and 1.6\% over the Max Rate method in terms of AP@0.5 and AP@0.7, respectively. Fig. \ref{fig_0.5} and Fig. \ref{fig_0.7} illustrate the AP performance of all methods as the bandwidth varies. The AP metric of all methods consistently increases as the bandwidth rises. The AP of the Max Features method is significantly lower than that of other methods. At 3 MHz, RACooper surpasses the Max Features method by 5.8\% in AP@0.5 and 8.2\% in AP@0.7. The reason is that the Max Features method focuses only on the two CAVs with the most remaining features during collaboration, overly emphasizing local information, which prevents the RSU’s from obtaining a comprehensive global view. The results of the Max Features method demonstrate the importance of prioritizing comprehensive perception information in collaborative perception. Compared to the Random method, RACooper achieves a 2.2\% improvement in AP@0.5 and a 6.1\% improvement in AP@0.7 at 3 MHz. This is because the Random method employs random subcarriers allocation and power control without considering the feature values of the collaborating vehicles, neglecting the feature complementarity with the RSU. 

Additionally, we compare detection loss ${L}_\text{det}$ and classification loss ${L}_\text{cls}$ across various bandwidth settings, as shown in Fig. \ref{fig_det_loss} and Fig. \ref{fig_cls_loss}. As the bandwidth increases, a clear reduction is observed in both  ${L}_\text{det}$ and ${L}_\text{cls}$. Owing to the direct optimization of the loss function, both the detection loss and classification loss of RACooper are lower than those of the other methods. An important observation is that although both ${L}_\text{det}$ and ${L}_\text{cls}$ decrease with increasing bandwidth, the classification loss ${L}_\text{cls}$ appears to more directly reflect the importance of optimizing the loss function for AP performance. This is evident from the fact that the Max Feature method achieves a lower detection loss ${L}_\text{det}$ than the Random method, yet exhibits a higher classification loss ${L}_\text{cls}$. Similarly, although its ${L}_\text{det}$ is close to that of the Max Rate method and RACooper, its ${L}_\text{cls}$ is markedly higher than all other methods. Together with the AP results shown in Fig. \ref{fig_ap}, the Max Feature method demonstrates considerably poorer performance than the other approaches. This indicates that, in the process of optimizing overall detection performance, we might consider increasing the weight of the classification loss.

From the combined results presented in Fig. \ref{fig_ap} and Fig. \ref{fig_loss}, it is evident that a lower detection loss tends to result in a higher AP, which also supports the assumption that minimizing detection loss indirectly optimizes the AP metric. The AP metric must be calculated sequentially for each object after every feature transmission in (\ref{equ_obr}). While optimizing the AP directly is computationally expensive and impractical for real-world collaborative perception scenarios, calculating the detection loss only involves the detection head for inference, which incurs negligible time overhead. As a result, RACooper is both efficient and suitable for deployment in practical collaborative perception scenarios.

The Max Rate method maximizes the sum rate of the V2I links, thus increasing the total number of features transmitted within the communication time. Considering the sparsity of the confidence map, most of the features transmitted by the Max Rate method are effective for the RSU, leading to a good performance in the AP metric. Although the Max Rate method maximizes the total number of transmitted features, our proposed method outperforms it by 1.0\% in AP@0.5 and 1.2\% in AP@0.7. This is because, as shown in (\ref{equ_s}), we consider not only the CSI but also the semantic importance of the features from collaborative vehicles. Consequently, RACooper learns a resource allocation policy based on perception value and channel state, resulting in better performance than the Max Rate method. Additionally, as shown in the comparison in  Fig. \ref{fig_0.5} and Fig. \ref{fig_0.7}, the performance gap between RACooper and other methods widens with the IoU threshold. The reason lies in the IoU threshold is positively correlated with localization accuracy, and our reward function includes the localization loss, leading to more accurate localization than other methods.

\begin{figure}[tb]
	\centering
	\includegraphics[width=0.85\linewidth]{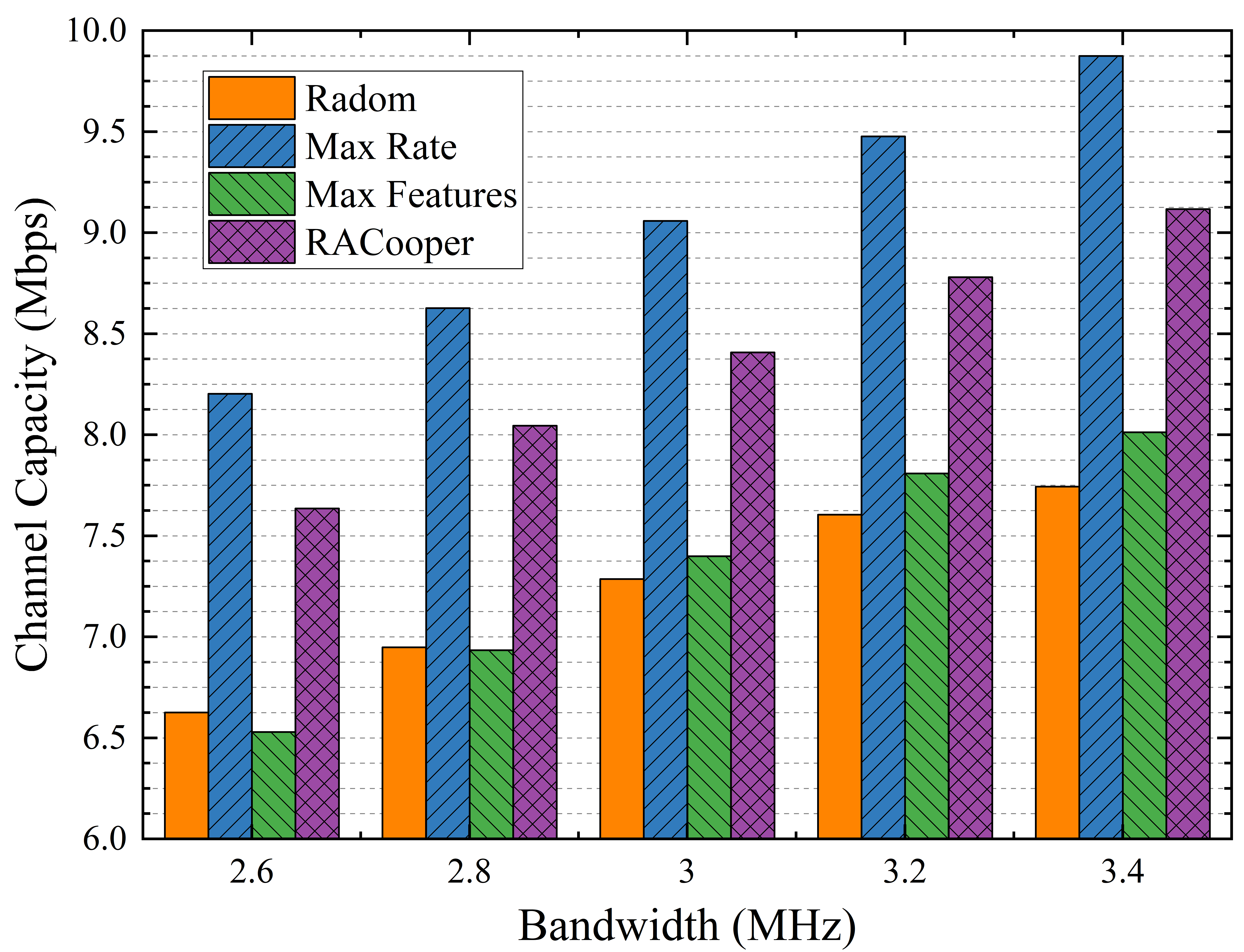}
	\caption{The channel capacity of our method compared to other baselines under different bandwidth conditions.}
	\label{fig_cap}
\end{figure}

\begin{figure}[t!]
	\centering
	\subfigure[AP@0.50 metric.]{
		\includegraphics[width=0.875\linewidth]{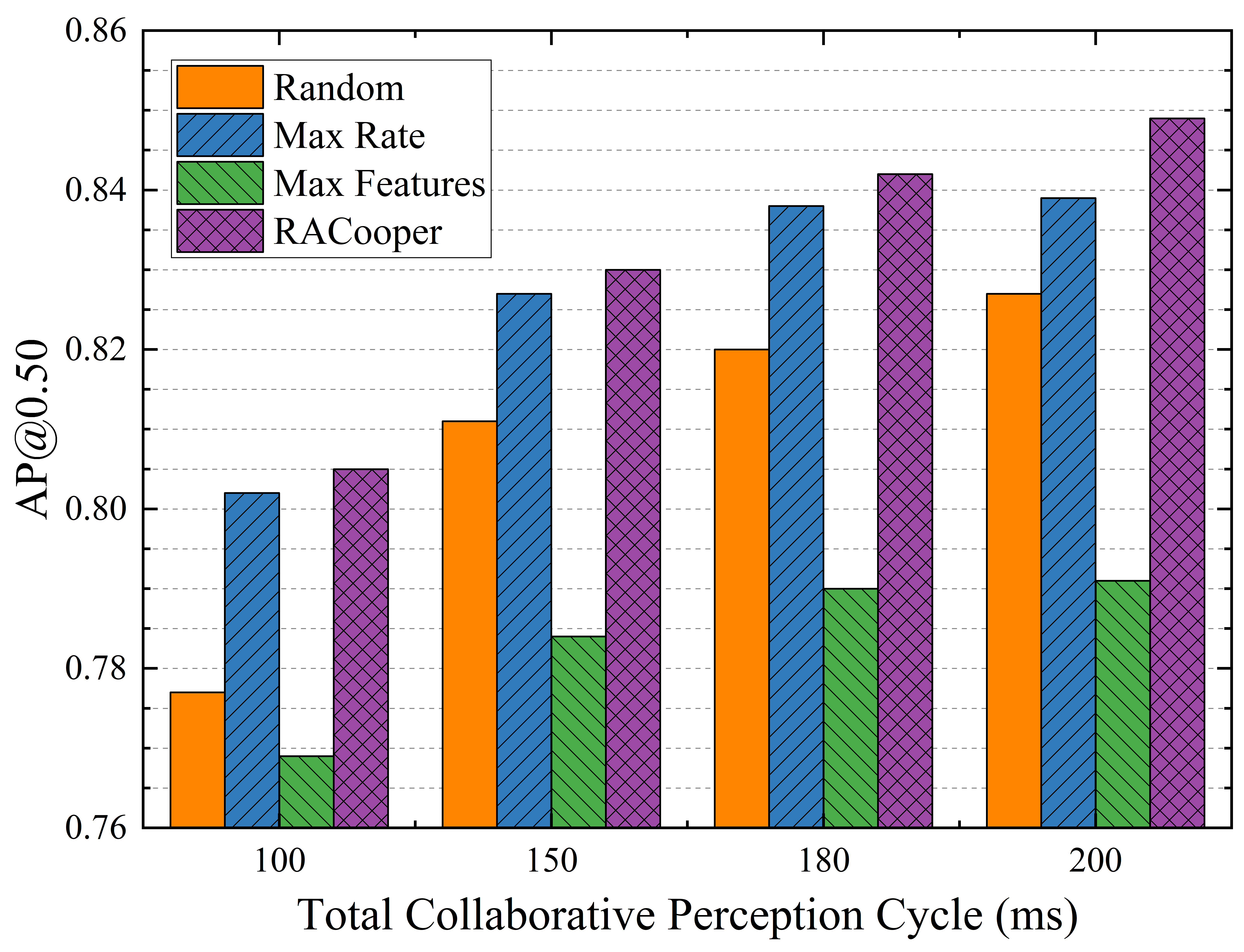}
		\label{fig_schet0.5} 
	}
	\subfigure[AP@0.70 metric.]{
		\includegraphics[width=0.875\linewidth]{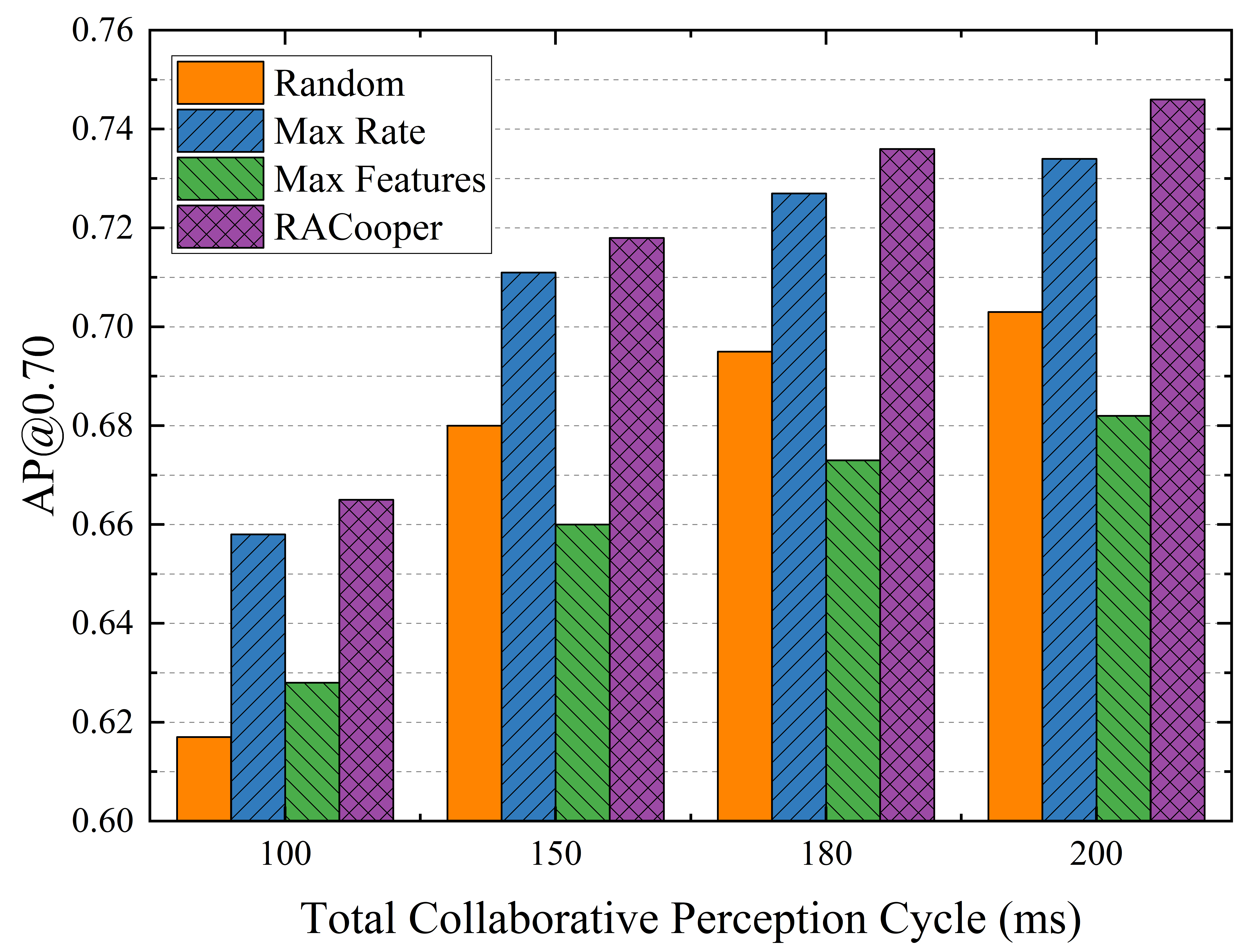}
		\label{fig_schet0.7} 
	}
	\DeclareGraphicsExtensions.
	\caption{The perception performance under different collaborative perception cycles in terms of AP@0.50 and AP@0.70.}
	\label{fig_schet}
\end{figure}

Fig. \ref{fig_cap} presents a further comparison of the channel capacities of various resource allocation methods at different bandwidths. Among these methods, the transmission rates of the Random method and Max Feature method are significantly lower than those of the others. The lower performance of the Random method is attributed to uncontrollable interference from random RBs and power allocation. For the Max Feature method, although there is no interference, only two collaborative vehicles transmit features at each step, limiting the channel capacity. As expected, the Max Rate method outperforms all other methods in transmission rate across all bandwidths. At 2.6 MHz, its transmission rate is 6.8\% higher than RACooper, and the gap widens as the bandwidth increases. Although RACooper lags behind the Max Rate method in communication rates, our method outperforms it in AP performance. In addition, the Max Rate method approach relies on exhaustive search to find the optimal solution, but this is constrained by limited computational resources and time in real-world scenarios, making it challenging to cope with the dynamic environment of C-V2X networks. As discussed above, RACooper achieves higher perception accuracy with a slight reduction in communication rates, improving collaborative perception performance while maintaining acceptable communication performance.

Fig. \ref{fig_schet} illustrates the AP@0.5 and AP@0.7 performance of each method under different collaborative perception cycles. Specifically, we analyze collaborative perception cycles of 100 ms, 150 ms, 180 ms, and 200 ms to simulate potential occlusion risks and severe communication congestion in real-world scenarios. In these scenarios, the RSU must collect as many relevant features as possible within shorter time intervals to ensure accurate perception. It can be observed that as the cycle increases, the AP@0.5 and AP@0.7 for all methods also improve. RACooper consistently outperforms the other methods across different collaborative cycles, even when the cycle is as short as 100 ms, achieving higher AP@0.5 and AP@0.7 values than the Max Rate method. Furthermore, RACooper surpasses the Max Rate method by 1.2\% and 1.6\% at 100 ms and 200 ms perception cycles, respectively. Although its performance slightly declines at 100 ms, the results still highlight RACooper’s robustness in scenarios with sudden occlusions and disrupted communication. The experimental results under different collaborative perception cycles indicate that RACooper prioritizes collaborative vehicles with higher perception value and better communication conditions, enabling it to obtain a better comprehensive view through complementary features in a very short time.

\subsection{Case Study: An Intersection Scenario}
\begin{figure}[t]
	\centering
	\includegraphics[width=0.8\linewidth]{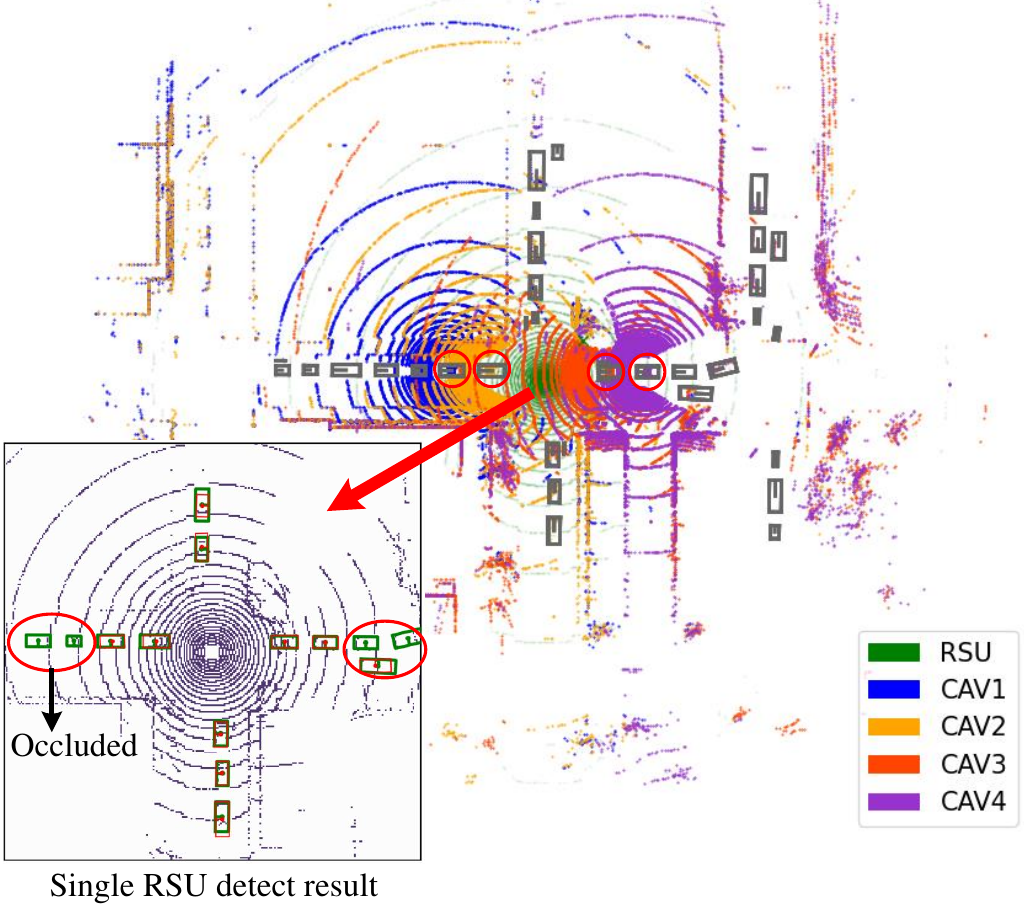}
	\caption{BEV of a collaborative perception scenario at an intersection. In the detection results of the single RSU, green boxes represent the ground truth, while red boxes indicate the predicted results.}
	\label{fig_bevo}
\end{figure}
Additionally, we simulate an intersection scenario to more clearly demonstrate the collaborative perception gains of each method in complex environments. As shown in Fig. \ref{fig_bevo}, the RSU is positioned at the center of the intersection, with four collaborative vehicles on either side, moving from right to left in the traffic flow. The resource allocation methods mentioned above are deployed on the RSU to collect the features of each CAVs, maximizing RSU's perception performance. Due to sensor range limitations and occlusion by surrounding vehicles, the RSU cannot perceive the four distant vehicles, limiting its perception performance.

\begin{figure}[t]
	\centering
	\subfigure[Random]{\includegraphics[width=0.175\textwidth]{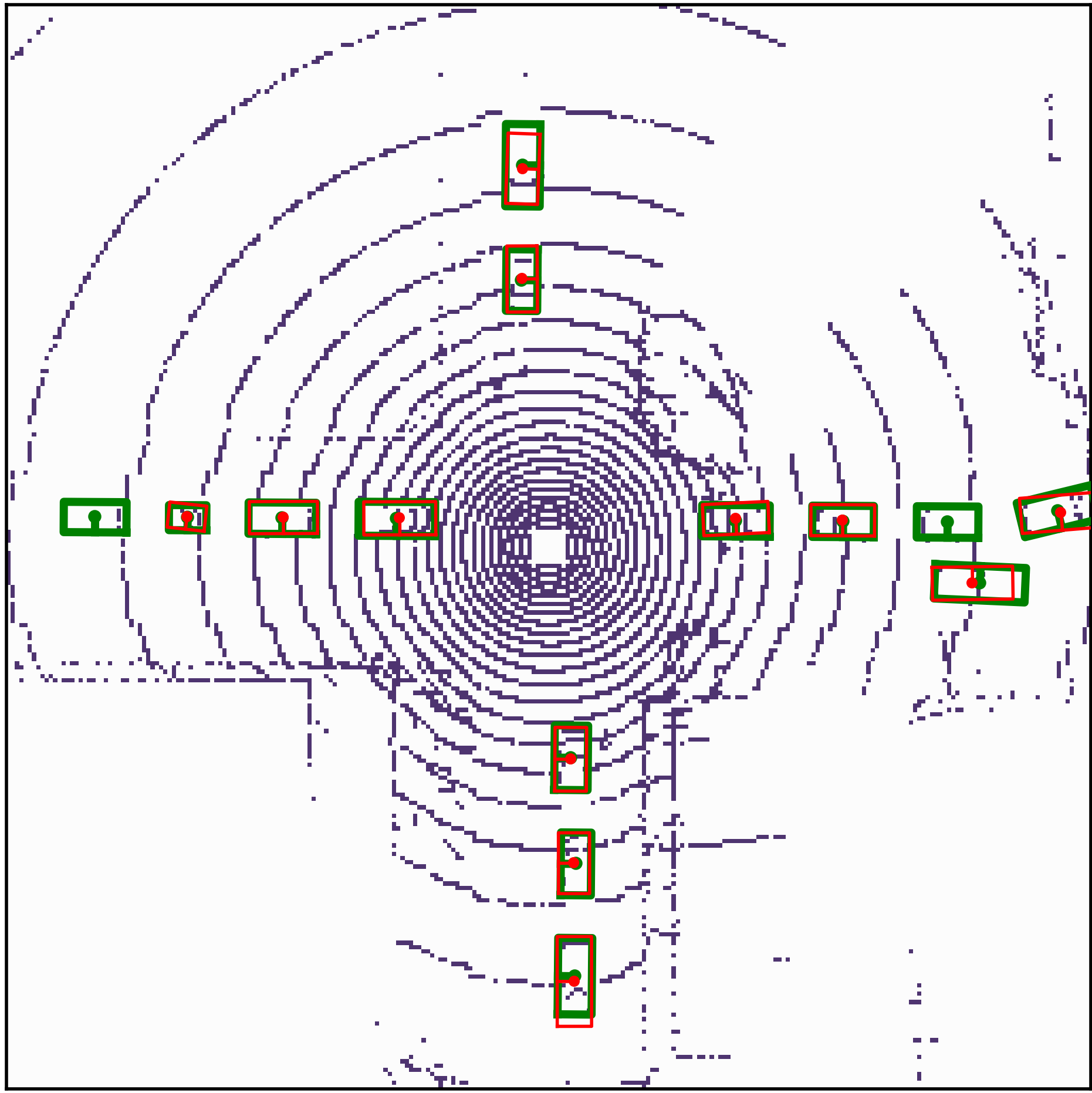}}
	\subfigure[Max Features]{\includegraphics[width=0.175\textwidth]{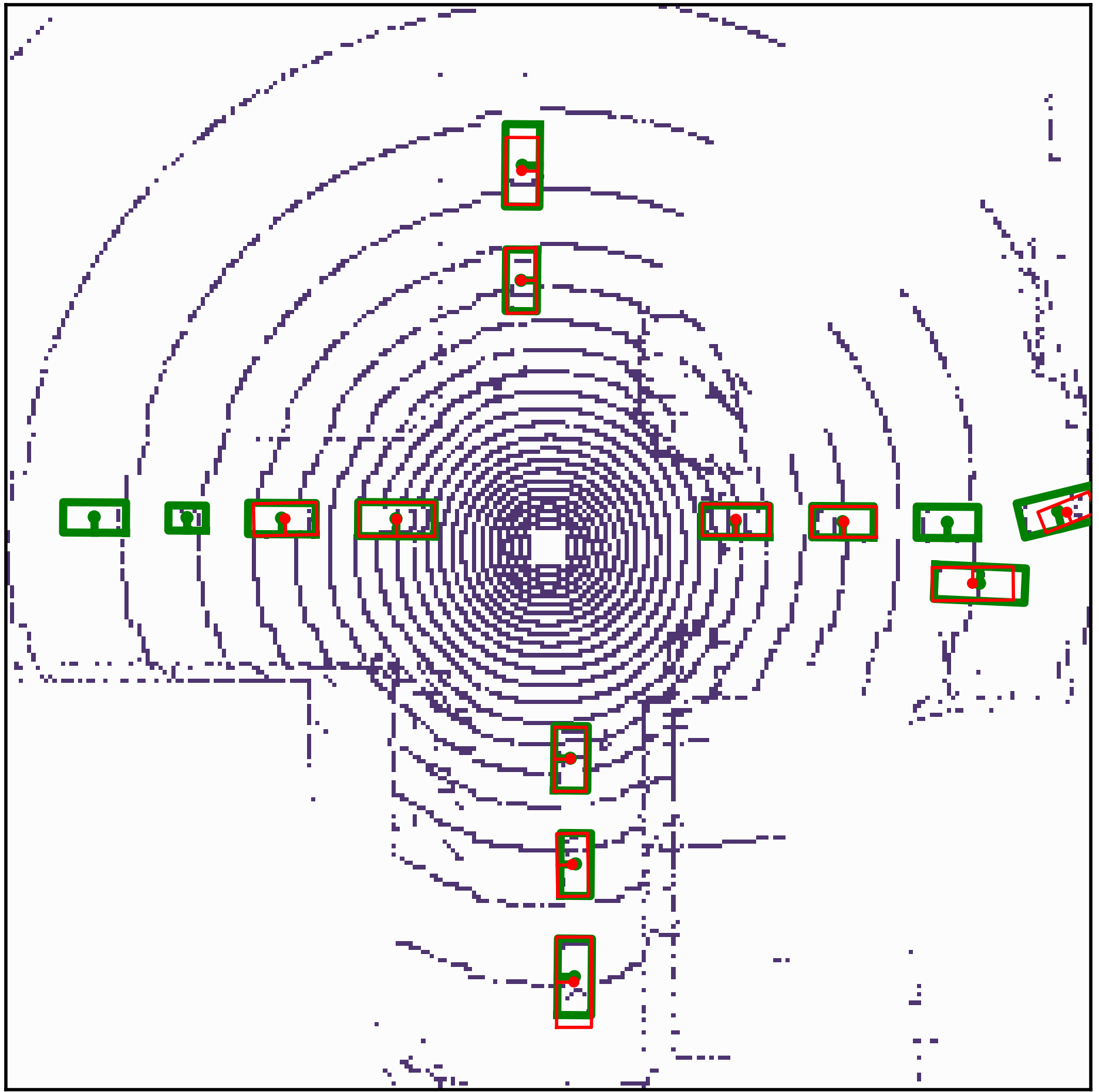}}
		
	\subfigure[Max Rate]{\includegraphics[width=0.175\textwidth]{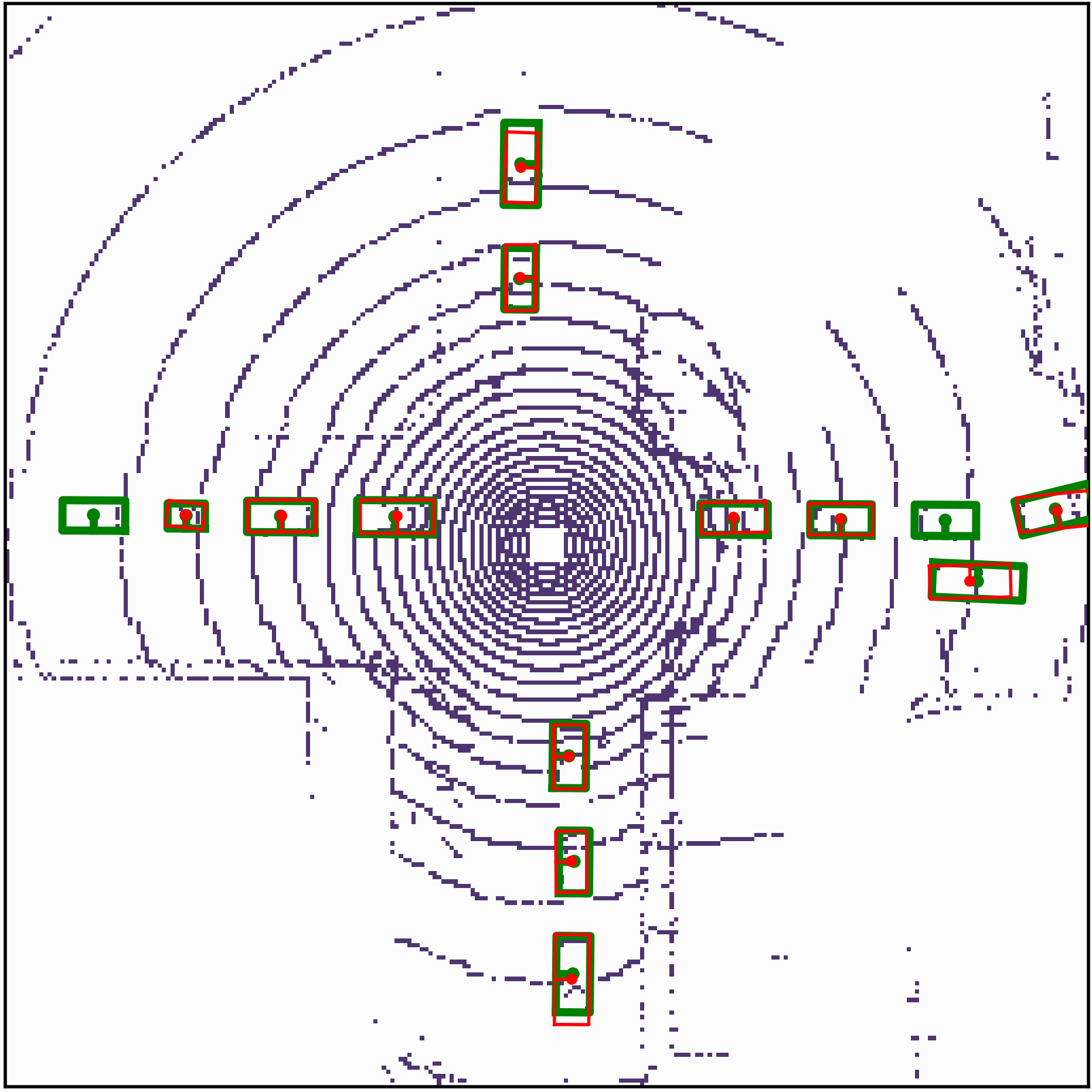}}
	\subfigure[RACooper]{\includegraphics[width=0.175\textwidth]{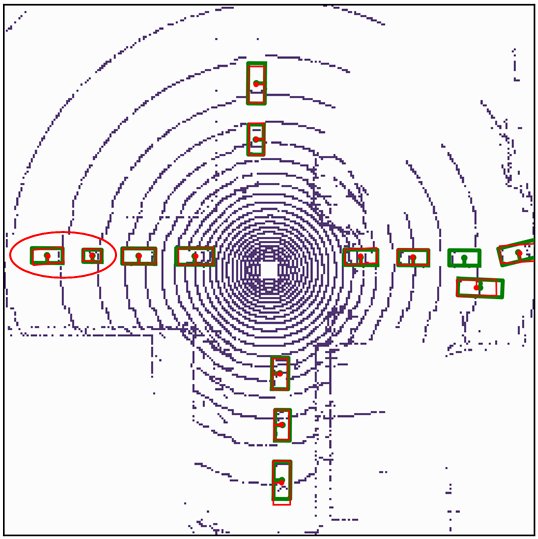}}
	
	\caption{Comparison of the detection result of different methods.}
	\label{fig_vis}
\end{figure}

\begin{figure}[t]
	\centering
	\subfigure[Communication rates]{
		\includegraphics[width=0.85\linewidth]{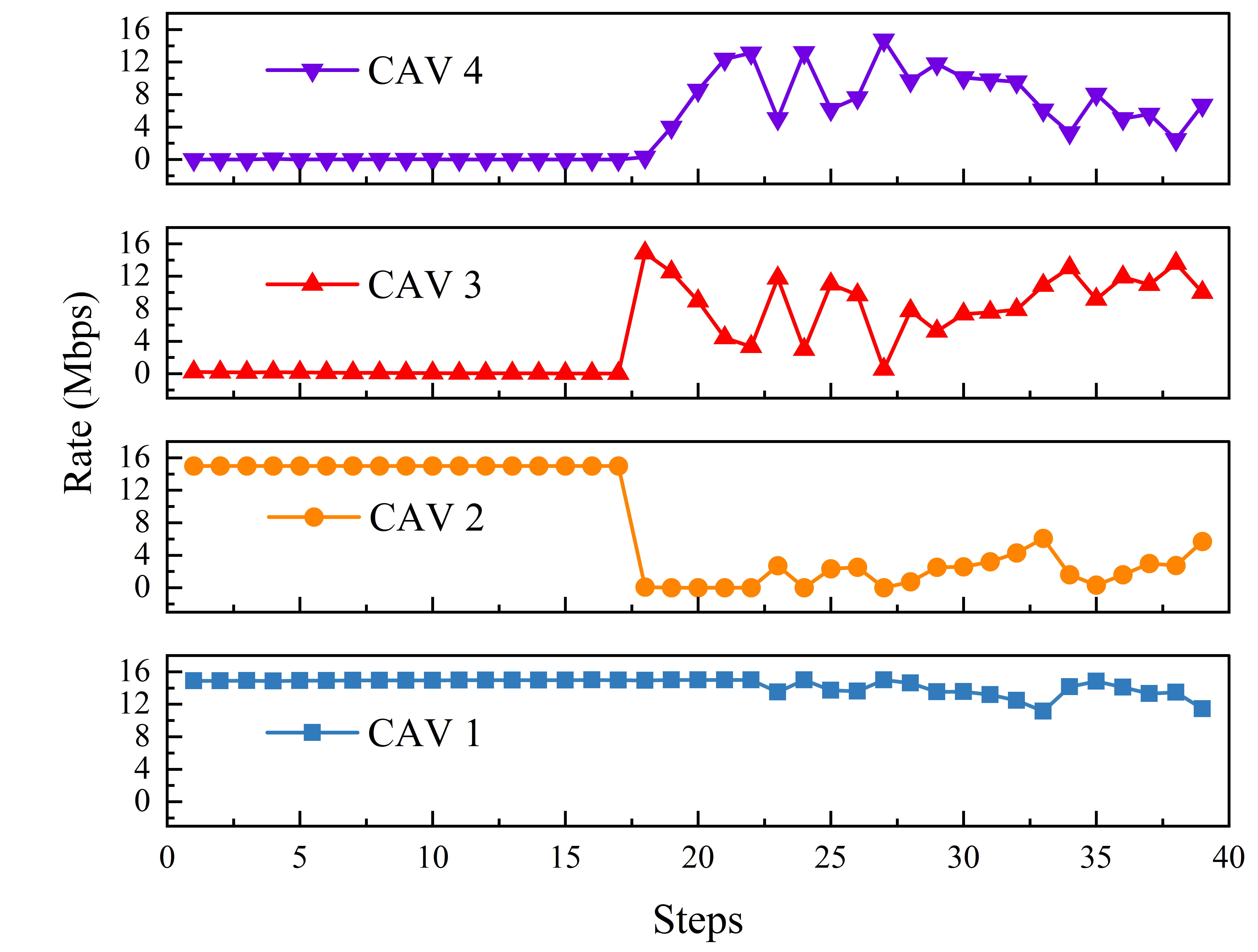}
		\label{fig_rate} 
	}
	\subfigure[Total confidence score]{
		\includegraphics[width=0.825\linewidth]{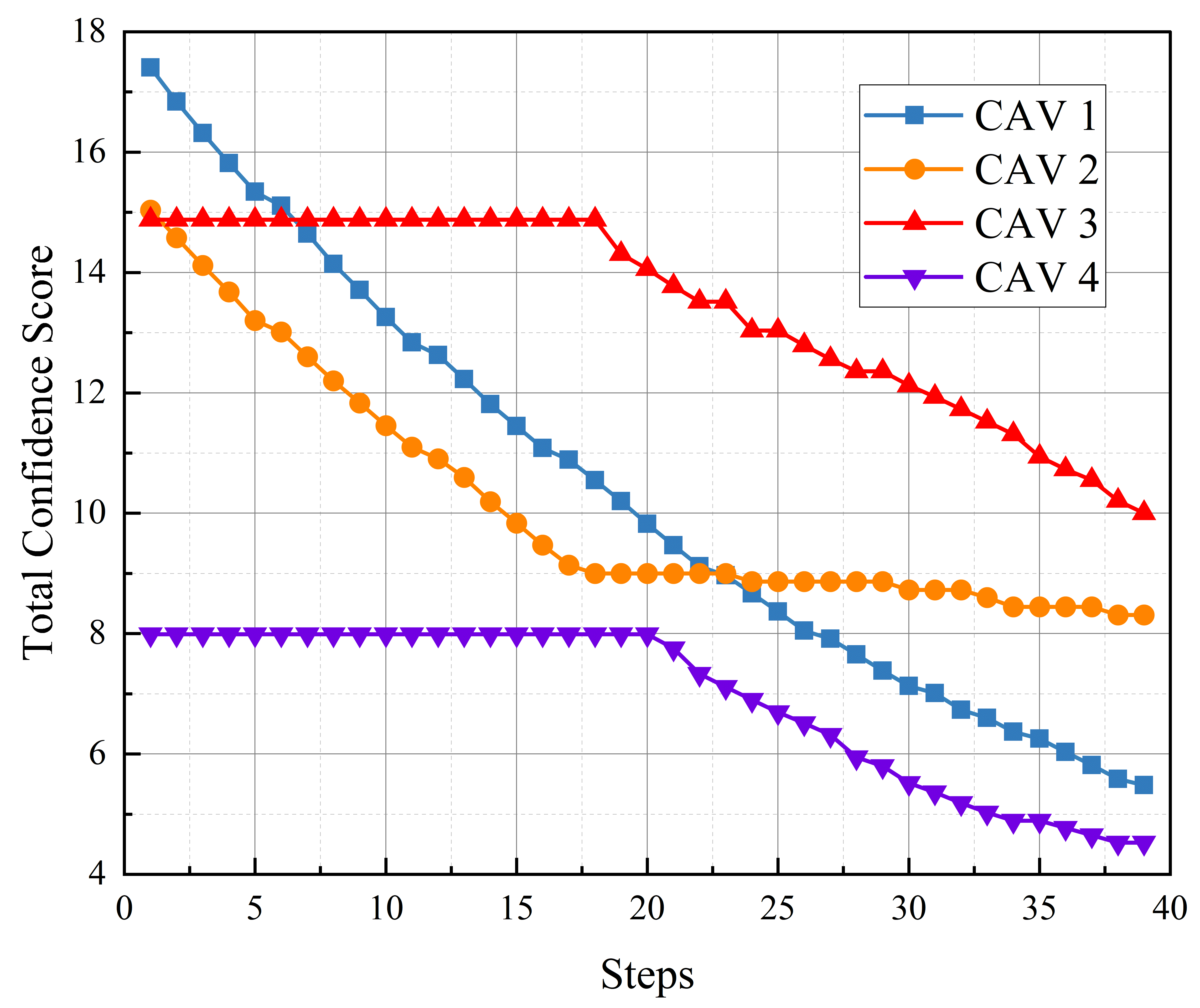}
		\label{fig_conf} 
	}
	\DeclareGraphicsExtensions.
	\caption{The total confidence score and communication rates for each CAV across time steps.}
	\label{fig_conf_rate}
\end{figure}

We evaluate the performance improvement of collaborative perception at the RSU for each method in this scenario. As shown in Fig. \ref{fig_vis}, the perception results for each resource allocation method are presented. A comparison with other methods reveals that none of the baseline methods are able to detect the target vehicles on the left side of the road, not even the Max Rate method. In contrast, RACooper, with a well-learned resource allocation strategy, obtains more effective perception information from the collaborative vehicles, successfully detecting all the vehicles on the left side of the road and accurately locating their positions, thus achieving better detection performance than the other baseline methods.

Meanwhile, we analyze the variations in each CAV’s communication rates and the overall confidence score, derived from $\mathbf{C}_m^{t}$ in (\ref{c_mt}), across the resource allocation time steps, as shown in Fig. \ref{fig_rate} and Fig. \ref{fig_conf}. It is evident that, in contrast to CAV 1 and CAV 2, CAV 4 has a lower overall confidence score as a result of heavier occlusion in its surroundings. Additionally, owing to the limited overlap between its perception area and that of the RSU, CAV 4 was not assigned resources for feature transmission in the early stages. From the beginning up to the 20th time step, feature transmission is primarily carried out by CAV 1 and CAV 2, indicating that they possess the most valuable information for the RSU. Fig. \ref{fig_bevo} shows that, due to the traffic flow predominantly moving from right to left and the limitations of LiDAR sensing range, CAV 1 and CAV 2 have limited perception of vehicles occluded behind them, unlike CAV 3 and CAV 4. As a result, their contribution to the RSU’s perception is less significant. Thus, under a perception gain driven resource allocation strategy, the RSU prioritizes feature transmission from CAV 1 and CAV 2 to compensate for the missing information of occluded objects. After 20th time step, the improvement in perception performance brought by a high communication rates reaches a bottleneck. When maintaining a high communication rates no longer results in significant perception gains, the RSU reduces the features uploaded by CAV 2 and reallocates the freed resources to CAV 3 and CAV 4. This allow them to upload features that complete the perception of regions behind them, which the RSU cannot sense on its own.

In summary, limited communication resources prevent the CAVs in this scenario from transmitting all features within a single collaborative perception period. A key challenge remains in determining how to transmit the most informative features. From the RSU’s perspective, CAV 1 provides the most effective perception of the occluded regions, prompting consistent resource allocation to CAV 1 across all time steps to ensure reliable feature transmission. This demonstrates that RACooper efficiently allocates more communication resources to high value transmission links by leveraging the complementarity between RSU and CAV features, thereby ensuring the transmission of critical information. Once the RSU collects sufficient features, its overall perception performance improves significantly. Building on this, RACooper provides CAVs with more comprehensive and robust perception outputs to support downstream tasks, thereby enhancing the safety of autonomous vehicles and other traffic participants.

\section{Conclusion}
In this paper, we propose a novel resource allocation framework, the RACooper, which integrates perception information and communication resources in collaborative perception, aiming to maximize the global perception accuracy for road decision-makers. To achieve this, we model the collaborative perception resource allocation problem under the V2I link through intermediate collaboration. To ensure computational efficiency and robustness, we transform the optimization problem from directly optimizing AP to minimizing the loss function, improving the stability of the RACooper training process. Considering that the optimization problem is NP-hard, we adopt advanced DRL methods to learn the resource allocation strategy and used a hierarchical DRL model to address the issue of large action space dimensions. Through extensive experiments, we demonstrate that the RACooper outperforms traditional resource allocation algorithms, achieving a balance between communication rates and perception performance, while also providing higher gains for collaborative perception in complex traffic scenarios.

\bibliographystyle{IEEEtran}
\bibliography{references.bib}

\begin{thebibliography}{10}
\providecommand{\url}[1]{#1}
\csname url@samestyle\endcsname
\providecommand{\newblock}{\relax}
\providecommand{\bibinfo}[2]{#2}
\providecommand{\BIBentrySTDinterwordspacing}{\spaceskip=0pt\relax}
\providecommand{\BIBentryALTinterwordstretchfactor}{4}
\providecommand{\BIBentryALTinterwordspacing}{\spaceskip=\fontdimen2\font plus
\BIBentryALTinterwordstretchfactor\fontdimen3\font minus
  \fontdimen4\font\relax}
\providecommand{\BIBforeignlanguage}[2]{{%
\expandafter\ifx\csname l@#1\endcsname\relax
\typeout{** WARNING: IEEEtran.bst: No hyphenation pattern has been}%
\typeout{** loaded for the language `#1'. Using the pattern for}%
\typeout{** the default language instead.}%
\else
\language=\csname l@#1\endcsname
\fi
#2}}
\providecommand{\BIBdecl}{\relax}
\BIBdecl

\bibitem{ren2022collaborative}
S.~Ren, S.~Chen, and W.~Zhang, ``Collaborative perception for autonomous
  driving: {Current} status and future trend,'' in \emph{Proc. CCSICC}, 2022,
  pp. 682--692.

\bibitem{li2022v2x}
Y.~Li, D.~Ma, Z.~An, Z.~Wang, Y.~Zhong, S.~Chen, and C.~Feng, ``{V2X-Sim:
  Multi-agent collaborative perception dataset and benchmark for autonomous
  driving},'' \emph{IEEE Robot. Autom. Lett.}, vol.~7, no.~4, pp.
  10\,914--10\,921, Oct. 2022.

\bibitem{noor2020survey}
M.~Noor-A-Rahim, Z.~Liu, H.~Lee, G.~M.~N. Ali, D.~Pesch, and P.~Xiao, ``A
  survey on resource allocation in vehicular networks,'' \emph{IEEE Trans.
  Intell. Transp. Syst}, vol.~23, no.~2, pp. 701--721, Sep. 2022.

\bibitem{gholmieh2021c}
R.~Gholmieh and H.~I. Abbasi, ``{C-V2X (LTE-V2X) performance enhancement
  through SAE J3161/1 probabilistic one-shot transmissions},'' in \emph{Proc.
  GLOBECOM}.\hskip 1em plus 0.5em minus 0.4em\relax IEEE, Dec. 2021, pp. 1--6.

\bibitem{chen2019cooper}
Q.~Chen, S.~Tang, Q.~Yang, and S.~Fu, ``Cooper: {Cooperative} perception for
  connected autonomous vehicles based on {3D} point clouds,'' in \emph{Proc.
  ICDCS}, 2019, pp. 514--524.

\bibitem{arnold2020cooperative}
E.~Arnold, M.~Dianati, R.~de~Temple, and S.~Fallah, ``{Cooperative perception
  for 3D object detection in driving scenarios using infrastructure sensors},''
  \emph{IEEE Trans. Intell. Transp. Syst.}, vol.~23, no.~3, pp. 1852--1864,
  Oct. 2020.

\bibitem{cao2025task}
Z.~Cao, H.~Zhang, L.~Liang, H.~Wang, S.~Jin, and G.~Y. Li, ``Task-oriented
  semantic communication for stereo-vision {3D} object detection,'' \emph{IEEE
  Trans. Commun.}, 2025.

\bibitem{miller2020cooperative}
A.~Miller, K.~Rim, P.~Chopra, P.~Kelkar, and M.~Likhachev, ``Cooperative
  perception and localization for cooperative driving,'' in \emph{Proc. ICRA},
  2020, pp. 1256--1262.

\bibitem{song2023cooperative}
Z.~Song, F.~Wen, H.~Zhang, and J.~Li, ``A cooperative perception system robust
  to localization errors,'' in \emph{Proc. IV}, 2023, pp. 1--6.

\bibitem{yu2022dair}
H.~Yu, Y.~Luo, M.~Shu, Y.~Huo, Z.~Yang, Y.~Shi, Z.~Guo, H.~Li, X.~Hu, J.~Yuan
  \emph{et~al.}, ``{Dair-V2X: A large-scale dataset for vehicle-infrastructure
  cooperative 3D object detection},'' in \emph{Proc. CVPR}, 2022, pp.
  21\,361--21\,370.

\bibitem{chen2019f}
Q.~Chen, X.~Ma, S.~Tang, J.~Guo, Q.~Yang, and S.~Fu, ``F-cooper: {Feature}
  based cooperative perception for autonomous vehicle edge computing system
  using {3D} point clouds,'' in \emph{Proc. SEC}, 2019, pp. 88--100.

\bibitem{guo2021coff}
J.~Guo, D.~Carrillo, S.~Tang, Q.~Chen, Q.~Yang, S.~Fu, X.~Wang, N.~Wang, and
  P.~Palacharla, ``{CoFF: Cooperative spatial feature fusion for 3-D object
  detection on autonomous vehicles},'' \emph{IEEE Internet Things J.}, vol.~8,
  no.~14, pp. 11\,078--11\,087, Jan. 2021.

\bibitem{abdel2021v2v}
M.~K. Abdel-Aziz, C.~Perlecto, S.~Samarakoon, and M.~Bennis, ``{V2V cooperative
  sensing using reinforcement learning with action branching},'' in \emph{Proc.
  ICC}, Aug. 2021, pp. 1--6.

\bibitem{zhou2022multi}
Y.~Zhou, J.~Xiao, Y.~Zhou, and G.~Loianno, ``Multi-robot collaborative
  perception with graph neural networks,'' \emph{IEEE Robot. Autom. Lett.},
  vol.~7, no.~2, pp. 2289--2296, Jan. 2022.

\bibitem{xu2022opv2v}
R.~Xu, H.~Xiang, X.~Xia, X.~Han, J.~Li, and J.~Ma, ``{OPV2V: An open benchmark
  dataset and fusion pipeline for perception with vehicle-to-vehicle
  communication},'' in \emph{Proc. ICRA}.\hskip 1em plus 0.5em minus
  0.4em\relax IEEE, 2022, pp. 2583--2589.

\bibitem{xu2022v2x}
R.~Xu, H.~Xiang, Z.~Tu, X.~Xia, M.-H. Yang, and J.~Ma, ``{V2X-ViT:
  Vehicle-to-everything cooperative perception with vision transformer},'' in
  \emph{Proc. ECCV}.\hskip 1em plus 0.5em minus 0.4em\relax Springer, 2022, pp.
  107--124.

\bibitem{hu2022where2comm}
Y.~Hu, S.~Fang, Z.~Lei, Y.~Zhong, and S.~Chen, ``{Where2comm:
  Communication-efficient collaborative perception via spatial confidence
  maps},'' in \emph{Proc. NIPS}, 2022, pp. 4874--4886.

\bibitem{wang2023vimi}
Z.~Wang, S.~Fan, X.~Huo, T.~Xu, Y.~Wang, J.~Liu, Y.~Chen, and Y.-Q. Zhang,
  ``{VIMI: Vehicle-infrastructure multi-view intermediate fusion for
  camera-based 3D object detection},'' \emph{arXiv preprint arXiv:2303.10975},
  2023.

\bibitem{liu2025deep}
Y.~Liu, G.~Liu, L.~Liang, H.~Ye, C.~Guo, and S.~Jin, ``Deep reinforcement
  learning-based user scheduling for collaborative perception,'' \emph{IEEE
  Trans. Mobile Comput.}, early access, Jul. 2025.

\bibitem{sheng2024semantic1}
Y.~Sheng, H.~Ye, L.~Liang, S.~Jin, and G.~Y. Li, ``Semantic communication for
  cooperative perception based on importance map,'' \emph{J. Franklin Inst.},
  vol. 361, no.~6, p. 106739, Apr. 2024.

\bibitem{clancy2024wireless}
J.~Clancy, D.~Mullins, B.~Deegan, J.~Horgan, E.~Ward, C.~Eising, P.~Denny,
  E.~Jones, and M.~Glavin, ``Wireless access for {V2X} communications:
  Research, challenges and opportunities,'' \emph{IEEE Commun. Surveys Tuts.},
  Apr. 2024.

\bibitem{machardy2018v2x}
Z.~MacHardy, A.~Khan, K.~Obana, and S.~Iwashina, ``{V2X access technologies:
  Regulation, research, and remaining challenges},'' \emph{IEEE Commun. Surveys
  Tuts.}, vol.~20, no.~3, pp. 1858--1877, Feb. 2018.

\bibitem{5gaa2021}
{5GAA Automotive Association}, ``Position paper on deployment band
  configuration for {C-V2X} at 5.9 {GHz} in {Europe},'' Jun. 2021.

\bibitem{chang2024interoperable}
J.~Chang, S.~G. Hatcher, E.~Kuruvilla, K.~Bare, M.~Zaatari, C.~Kain,
  I.~McManus, A.~Seshadri, K.~Thompson \emph{et~al.}, ``Interoperable
  connectivity—national {V2X} deployment plan supplement,'' United States.
  Department of Transportation., Tech. Rep., 2024.

\bibitem{chan2024rsu}
M.~D.~T. Chan, Z.~Nan, Y.~Jia, S.~Zhou, and Z.~Niu, ``Rsu-aided
  energy-efficient collaborative perception for connected autonomous
  vehicles,'' in \emph{Proc. WCNC}.\hskip 1em plus 0.5em minus 0.4em\relax
  IEEE, 2024, pp. 1--6.

\bibitem{abdel2021vehicular}
M.~K. Abdel-Aziz, C.~Perfecto, S.~Samarakoon, M.~Bennis, and W.~Saad,
  ``Vehicular cooperative perception through action branching and federated
  reinforcement learning,'' \emph{IEEE Trans. Commun.}, vol.~70, no.~2, pp.
  891--903, Nov. 2021.

\bibitem{luo2023edgecooper}
G.~Luo, C.~Shao, N.~Cheng, H.~Zhou, H.~Zhang, Q.~Yuan, and J.~Li,
  ``{Edgecooper: Network-aware cooperative lidar perception for enhanced
  vehicular awareness},'' \emph{IEEE J. Sel. Areas Commun.}, Oct. 2023.

\bibitem{ye2023accuracy}
X.~Ye, K.~Qu, W.~Zhuang, and X.~Shen, ``Accuracy-aware cooperative sensing and
  computing for connected autonomous vehicles,'' \emph{IEEE Trans. Mobile
  Comput.}, vol.~23, no.~8, pp. 8193--8207, Dec. 2023.

\bibitem{jia2024c}
Y.~Jia, Y.~Sun, R.~Mao, Z.~Nan, S.~Zhou, and Z.~Niu, ``{C-MASS}: Combinatorial
  mobility-aware sensor scheduling for collaborative perception with
  second-order topology approximation,'' \emph{arXiv preprint
  arXiv:2407.00412}, 2024.

\bibitem{sheng2024semantic}
Y.~Sheng, L.~Liang, H.~Ye, S.~Jin, and G.~Y. Li, ``Semantic communication for
  cooperative perception using {HARQ},'' \emph{IEEE Trans. Cogn. Commun.
  Netw.}, early access, Jun. 2025.

\bibitem{molina2017lte}
R.~Molina-Masegosa and J.~Gozalvez, ``{LTE-V for sidelink 5G V2X vehicular
  communications: A new 5G technology for short-range vehicle-to-everything
  communications},'' \emph{IEEE Veh. Technol. Mag.}, vol.~12, no.~4, pp.
  30--39, Dec. 2017.

\bibitem{lang2019pointpillars}
A.~H. Lang, S.~Vora, H.~Caesar, L.~Zhou, J.~Yang, and O.~Beijbom,
  ``Pointpillars: {Fast} encoders for object detection from point clouds,'' in
  \emph{Proc. ICCV}, 2019, pp. 12\,697--12\,705.

\bibitem{zhao2023adaptive}
J.~Zhao, H.~Quan, M.~Xia, and D.~Wang, ``Adaptive resource allocation for
  mobile edge computing in internet of vehicles: {A} deep reinforcement
  learning approach,'' \emph{IEEE Trans. Veh. Technol.}, vol.~73, no.~4, pp.
  5834--5848, Apr. 2023.

\bibitem{an2024channel}
H.~An, Z.~Fang, Y.~Zhang, S.~Hu, X.~Chen, G.~Xu, and Y.~Fang, ``Channel-aware
  throughput maximization for cooperative data fusion in {CAV},'' \emph{arXiv
  preprint arXiv:2410.04320}, 2024.

\bibitem{schulman2017proximal}
J.~Schulman, F.~Wolski, P.~Dhariwal, A.~Radford, and O.~Klimov, ``Proximal
  policy optimization algorithms,'' \emph{arXiv preprint arXiv:1707.06347},
  2017.

\bibitem{3gpp2018nr}
3rd Generation Partnership~Project, ``{Technical specification group radio
  access network; NR; study on NR vehicle-to-everything(V2X); (Release 16)},''
  Mar. 2018.

\end{thebibliography}

\newpage

\vfill

\end{document}